\documentclass[titlepage,12pt]{article}
\usepackage{amssymb,amsmath,amsfonts}
\usepackage[utf8]{inputenc}
\usepackage{setspace}
\usepackage{epsfig}
\usepackage{graphicx,graphics}
\usepackage{verbatim}
\usepackage{cite}
\usepackage{slashed}
\usepackage{textcomp}

\usepackage[T1]{fontenc} 
\usepackage{multirow}
\usepackage{caption}

\usepackage{enumitem}

\makeatletter
\def\@sect#1#2#3#4#5#6[#7]#8{\ifnum #2>\c@secnumdepth
  \def\@svsec{}\else
  \refstepcounter{#1}\edef\@svsec{\csname the#1\endcsname.\hskip0.5em}\fi
  \@tempskipa #5\relax
  \ifdim \@tempskipa>\z@
    \begingroup
      #6\relax
      \@hangfrom{\hskip #3\relax\@svsec}{\interlinepenalty \@M #8\par}%
    \endgroup
    \csname #1mark\endcsname{#7}\addcontentsline
      {toc}{#1}{\ifnum #2>\c@secnumdepth \else
        \protect\numberline{\csname the#1\endcsname}\fi #7}%
  \else
    \def\@svsechd{#6\hskip #3\@svsec #8\csname #1mark\endcsname
      {#7}\addcontentsline{toc}{#1}{\ifnum #2>\c@secnumdepth \else
        \protect\numberline{\csname the#1\endcsname}\fi #7}}%
  \fi \@xsect{#5}}

\newcommand{\ttbar}{t{\bar t}}
\newcommand{\mtt}{M_{t{\bar t}}}
\newcommand{\mWW}{M_{WW}}

\begin{document}
\begin{titlepage}
  \begin{flushright}
TTK-15-35
    \end{flushright}
\vspace{0.01cm}
\begin{center}
{\LARGE \bf Improved effective vector boson approximation revisited} \\
\vspace{1.5cm}
{\bf Werner Bernreuther}\,\footnote{\tt breuther@physik.rwth-aachen.de} {\bf and  Long Chen}\,\footnote{\tt algeochen@physik.rwth-aachen.de}
\par\vspace{1cm}
Institut f\"ur Theoretische Teilchenphysik und Kosmologie, RWTH Aachen University, \\ 52056 Aachen, Germany
\par\vspace{1cm}
{\bf Abstract}\\
\parbox[t]{\textwidth}
{\small{ We reexamine the improved effective vector boson approximation which is based on two-vector-boson luminosities $\mathrm{\mathbf{L}}_{\rm pol}$ 
  for the computation of  weak gauge-boson hard scattering subprocesses $V_1 V_2\to {\cal W}$ in high-energy hadron-hadron or
   $e^-e^+$ collisions.  We calculate these luminosities for the nine combinations of the transverse and longitudinal polarizations 
   of $V_1$ and $V_2$ in the unitary and axial gauge. For these two gauge choices 
  the quality of this approach is investigated  for the reactions $e^-e^+ \to W^- W^+  \nu_e \bar{\nu}_e$ and $e^-e^+ \to \ttbar  \nu_e \bar{\nu}_e$ 
  using appropriate phase-space cuts. 
}}

\end{center}
\vspace*{0.7cm}

PACS number(s): 12.15.Ji, 12.38.Bx, 13.66.Fg 
\end{titlepage}
%
%
\setcounter{footnote}{0}
\renewcommand{\thefootnote}{\arabic{footnote}}
\setcounter{page}{1}

\section{Introduction} 
\label{sec:intro}

Although the discovery of the 125 GeV Higgs boson  \cite{Aad:2012tfa,Chatrchyan:2012xdj} at the  Large Hadron Collider 
strongly supports the Higgs mechanism of electroweak symmetry breaking (EWSB), it does not exclude the possibility
that additional (spin-zero) resonances linked to EWSB with masses  in or below the TeV range exist. 
Therefore, the detailed exploration of this issue remains to be 
one of the prime present and future research goals at this machine and 
at future high-energy proton-proton or electron-positron colliders that are presently being discussed.
One of the most direct probes of the dynamics of EWSB is the high-energy scattering 
of electroweak gauge bosons $V=W^\pm, Z$, especially of longitudinally polarized 
ones \cite{Chanowitz:1985hj,Bagger:1995mk,Englert:2008tn,Doroba:2012pd}. 
As weak gauge-boson beams are not available, $V_1 V_2$ scattering or fusion can be studied at $pp$ or $e^-e^+$
colliders only through reactions of the form
$f_1 f_2\to f'_1f'_2 {\cal W}$, where the $f_i,f'_i$ denote quarks (leptons) in the
case of $pp$ $(e^-e^+)$ colliders. Typical final states ${\cal W}$ 
of interest are a heavy non-standard Higgs boson, a weak gauge-boson
pair $V'_1 V'_2$, or a top-quark top antiquark $(t{\bar t})$ pair.    
At very high energies  such reactions, which involve the scattering or fusion of two vector bosons,
have often been analyzed by means of the effective vector boson approximation 
(EVBA) \cite{Cahn:1983ip,Dawson:1984gx,Kane:1984bb}. In this approximation the vector boson $V$ radiated
off a (anti)quark or electron/positron is treated as a constituent of the respective fermion.
 In the pioneering works  \cite{Cahn:1983ip,Dawson:1984gx,Kane:1984bb} the weak gauge boson distribution functions were computed in the
leading logarithmic approximation. The QCD radiative corrections to these functions were calculated in \cite{Dawson:1988ai}. 
The method was validated  in \cite{Kunszt:1987tk} within the axial gauge 
for the case of heavy Higgs-boson production \cite{Kunszt:1987tk} that is dominated by the fusion of two longitudinally
 polarized weak gauge bosons, and more recently in \cite{Borel:2012by} using the same gauge. 
  The applicability and limitations of the EVBA in the 
leading logarithmic approximation and of improved versions \cite{Johnson:1987tj,Kuss:1995yv}
to heavy fermion production and to $V_1 V_2\to V'_1 V'_2$ scattering  have been analyzed in many papers, including
\cite{Kunszt:1987tk,Kleiss:1986xp,Willenbrock:1986cr,Gunion:1986gm,Dawson:1986tc,Kauffman:1989aq,Larios:1997ey,Accomando:2006mc,Accomando:2006hq,Bouayed:2007rt,Alboteanu:2008my,Borel:2012by}.
To date one may question the need of this approximation, which singles out a certain class of contributions
to the complete scattering amplitude, especially in view that powerful computer packages exist, 
including \cite{Kilian:2007gr,Alwall:2014hca} at leading order and  \cite{Alwall:2014hca,Arnold:2011wj,Baglio:2014uba} at 
next-to-leading order, which allow to numerically compute the respective processes exactly at the respective 
order of perturbation theory. 
Yet, the EVBA may still be useful in appropriate kinematic regions
as a tool for analyzing in a transparent way weak gauge-boson reactions that are relevant for the physics 
of electroweak symmetry breaking; cf. for instance, the recent applications \cite{Borel:2012by,Brehmer:2014pka}.
      
A critical point of the  EVBA in the leading logarithmic expansion are the  approxi\-mations 
in the computation of the vector-boson distribution functions $F_\lambda(\xi)$.
(Here $F_\lambda(\xi)$ has the usual interpretation as the probability of finding a vector boson with helicity $\lambda$
and longitudinal momentum fraction $\xi$ in an incoming high-energy fermion $f$.)
While the leading logarithmic approximation works reasonably well for a longitudinally polarized vector boson
if $\xi>0.05$ and the center-of-mass energy of the initial state is larger than $\sim 1$ TeV, the 
distribution functions $F_T(\xi)$ for a transversely polarized weak gauge 
boson computed in the leading logarithmic expansion considerably overestimate the respective 
exact distribution functions \cite{Johnson:1987tj}. The distribution functions presented in 
\cite{Johnson:1987tj} were calculated without  approximations related to kinematics. 
A further improvement of the EVBA was worked out in \cite{Kuss:1995yv} for the case of two-vector boson 
processes, which are the reactions of interest for probing the dynamics of EWSB. Simple convolutions of 
two single vector boson distribution functions do not account for the mutual influence of the 
emission of boson $V_1$ on the probability for the emission of $V_2$ and vice versa.
This is incorporated in the two-vector-boson luminosities derived in \cite{Kuss:1995yv} in the unitary gauge. 
Moreover, non-diagonal terms in the summation over the polarizations
of $V_1$ and $V_2$ were also taken into account in this work, and no kinematic approximations were used. In this approach,
 a dynamical approximation remains, namely the on-mass-shell continuation of the $V_1 V_2\to {\cal W}$ 
hard scattering cross section. Yet the set of these correlated two-vector boson luminosities is gauge-dependent.

 The fact that the subset of diagrams to the reactions $f_1 f_2\to f'_1f'_2 {\cal W}$ which describe the scattering of two off-shell
 gauge bosons $V_1 V_2 \to {\cal W}$ is gauge dependent
  is another critical point of the (improved) EVBA.
 It is well-known that in particular in the 
  unitary gauge the off-shell hard scattering sub-amplitudes show, apart from specific examples, 
 a bad high-energy behavior  \cite{Kleiss:1986xp}. 
  It was argued in  \cite{Kunszt:1987tk,Borel:2012by} 
   that in the axial gauge and using the EVBA in the leading logarithmic approximation  the problem of bad off-shell 
    behavior can be avoided and the effective vector boson approximation works in this gauge if certain (kinematic) conditions are met. 
    In \cite{Accomando:2006mc} numerical studies of $W^+W^-$ production were performed by computing both the full set of
     Feynman diagrams and the subset of scattering diagrams associated with $W^+W^-\to W^+W^-$, using the unitary, axial, and a covariant gauge.
     It was found that when computing the cross section with the scattering diagrams only, the axial gauge (for a specific choice of the associated
     vector $n^\mu$) yields within these gauge-choices the best approximation to the full, gauge-independent cross section.
     Applications of the improved EVBA formulation of \cite{Kuss:1995yv}, which uses the dynamical approximation mentioned above (cf.
     Sec.~\ref{sec:LumiF}), include \cite{Accomando:2006mc,Accomando:2006hq,Kuss:1996ww},
     with conclusions that are not unanimous. While \cite{Accomando:2006mc} states that this framework provides not more than a very rough estimate,
     Ref.~\cite{Accomando:2006hq} and~\cite{Kuss:1996ww} report, for ${\cal W}=W^+W^-$ and ${\cal W}= Z Z$,
     an agreement of this approximation with the full result within about $20\%$ to $25\%$ and $10\%$, respectively.

     The effective vector boson approximation in the axial gauge using the EVBA in the leading logarithmic approximation was recently analyzed 
     in detail for single $W$-boson emission \cite{Borel:2012by}.
     One may ask whether the improved EVBA set-up with correlated 
     two-vector-boson luminosities \cite{Kuss:1995yv} derived in the axial gauge provides
      a useful approximation to processes that involve the scattering of 
      two gauge bosons. \\

In this paper we revisit the approach of \cite{Kuss:1995yv} which we call here the improved  effective 
vector boson approximation. We reexamine the two-vector-boson luminosities given in \cite{Kuss:1995yv} 
in the unitary gauge and clarify an issue related to relative minus signs. As a new aspect we compute
the two-vector-boson luminosities which involve a parity-odd combination of the vector 
and axial vector coupling of $V_1$ or $V_2$.  They are 
relevant for processes where the hard-scattering matrix element 
$V_1(\lambda_1) V_2(\lambda_2)\to {\cal W}$ involves parity-violating interactions. 
This is the case, for instance, for ${\cal W}=t{\bar t}$ in the Standard Model. 
 In addition we compute the two-vector-boson luminosities in the axial gauge. To our knowledge this is a new result.
Moreover, we investigate the quality of the improved EVBA in both gauges for two examples. 
To be specific we consider high-energy  $e^-e^+$ collisions and analyze the processes $e^-e^+\to W^-W^+\nu_e{\bar \nu_e}$
and $e^-e^+\to t {\bar  t} \nu_e{\bar \nu_e}$  to lowest order  in the Standard Model. 
We compute the respective tree-level cross sections both 
within the improved EVBA  and fully with the computer code MadGraph \cite{Alwall:2014hca}, i.e.,
taking into account all contributing Feynman diagrams, and compare the relative differences 
 using appropriate cuts on the final-state particles.

The paper is organized as follows. In Sec.~\ref{sec:LumiF} we outline the approach of Kuss and 
Spiesberger (KS) \cite{Kuss:1995yv} in deriving improved two-vector-boson luminosities. 
We clarify an issue related to relative minus signs and we compute 
luminosities which involve parity-even and  parity-odd combinations of the vector and axial 
vector coupling of $V_1$ or $V_2$.  The formulae apply to both incoming quarks and leptons.  
For $W^-W^+$ bosons radiated off $e^-e^+$, we compare our results
for the ``parity-even'' luminosities with those of \cite{Kuss:1995yv}. 
 In Sec.~\ref{sec:LumiAx} we compute the two-vector-boson luminosities in the axial gauge.
Sec.~\ref{sec:applic} contains our comparison of the cross sections for 
$e^-e^+\to W^-W^+\nu_e{\bar \nu_e}$ and  $e^-e^+\to t {\bar  t} \nu_e{\bar \nu_e}$ 
computed exactly and with the improved EVBA. 
We conclude in Sec.~\ref{sec:concl}. 
Appendices A and B contain our results for the four-fold differential  luminosities in the unitary and axial gauge.

\section{The KS luminosity formula}
\label{sec:LumiF}

We consider the production of an arbitrary state $\mathcal{W}$ by 
the scattering of two light fermions:
\begin{eqnarray}
f_1(l_1) + f_2(l_2) \rightarrow f'_1(l'_1) + f'_2(l'_2) + \mathcal{W}(p_\mathcal{W}) \, ,
\label{The2To3Process}
\end{eqnarray}
where $f_i$ $(f'_i)$ denote the fermions in the initial (final) state and the symbols in brackets 
are the associated four-momenta. The cross section of this process is given by
\begin{eqnarray}
\sigma_{f_1 f_2} 
&=&
\frac{1}{2 s}
\int \mathrm{d}\Gamma_2
~ \mathrm{d}{\tilde p}_{\mathcal{W}}~
\delta^{(4)}\left(l_1 + l_2 -l'_1 - l'_2 -p_ \mathcal{W} \right)
\overline{
|\mathcal{M}_{f_1 f_2 \rightarrow f'_1  f'_2 \mathcal{W}} |^2} \, ,
\label{TotalCS-starting}
\end{eqnarray}
where $s = \left( l_1 + l_2\right)^2$, $\mathrm{d}\Gamma_2\equiv 
 {\mathrm{d}^3 \vec{l}'_1}{\mathrm{d}^3 \vec{l}'_2}/{(16\pi^2 E'_1 E'_2)}$, and 
 $\mathrm{d}{\tilde p}_{\mathcal{W}}\equiv \mathrm{d}^3 p_{\mathcal{W}}/[(2\pi)^3 2 E_{\mathcal{W}}]$.
 Moreover, $\overline{|\mathcal{M}|^2}$ denotes the squared matrix element
 of \eqref{The2To3Process} which is averaged and summed over the helicities 
 (and colors, in the case of quarks) of the fermions $f_i$ and $f'_i$, respectively.

\begin{figure}[h!]
\centering
\includegraphics[width=5.5cm,height=5.5cm]{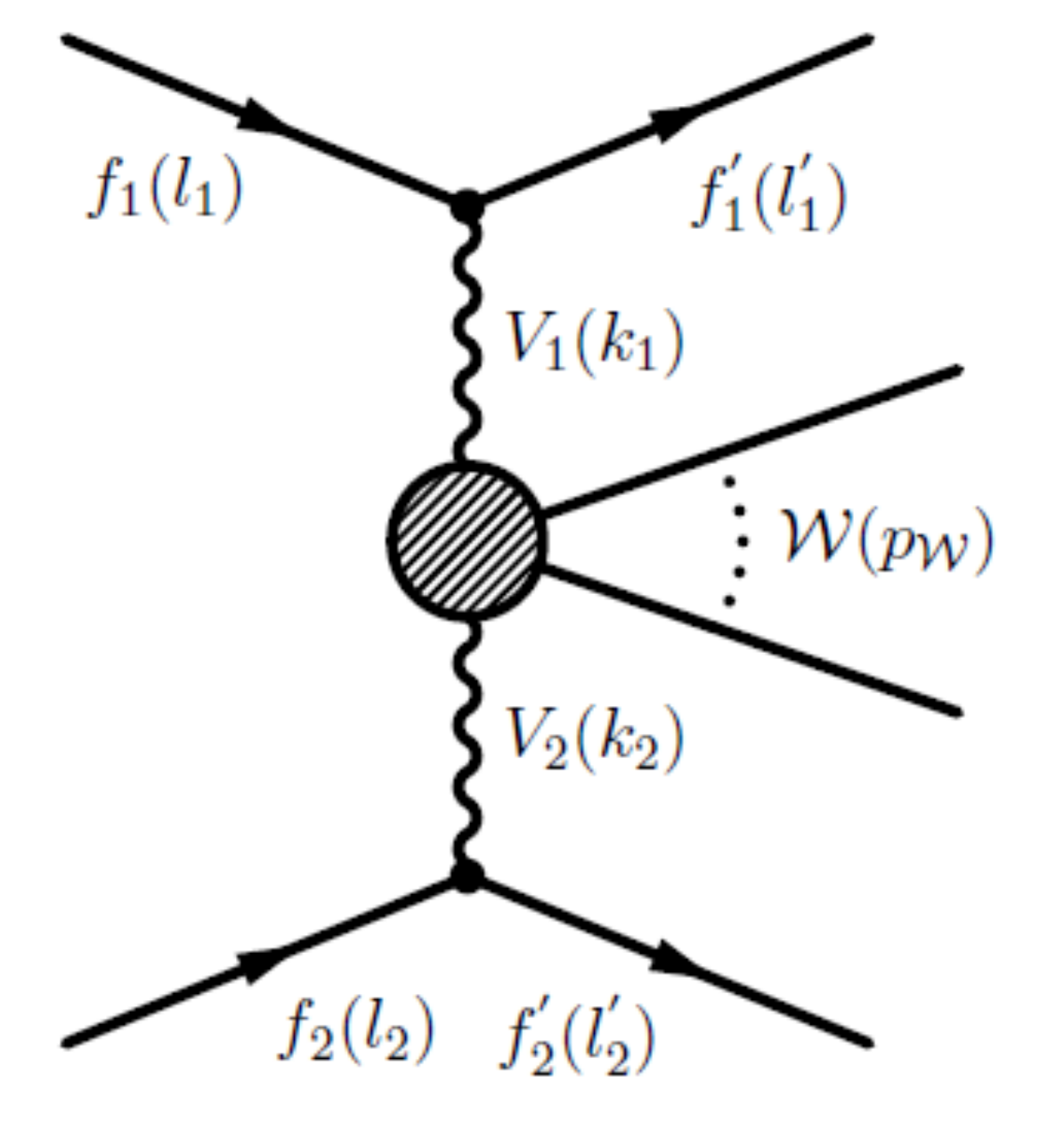} 
\caption{Generic weak gauge-boson scattering diagram.}         
\label{fig:WWscatt}
\end{figure}

In the following we consider processes \eqref{The2To3Process} which proceed via the exchange of
two off-shell weak gauge bosons $V_1, V_2$ $(V=W,Z)$ with masses $m_1, m_2$, 
as depicted in Fig.~\ref{fig:WWscatt}.

In the unitary gauge, which was used in \cite{Kuss:1995yv}, the  matrix element which corresponds to the diagram Fig.~\ref{fig:WWscatt}
 takes the form
\begin{eqnarray}
\mathcal{M}_{f_1 f_2 \rightarrow f'_1 f'_2 \mathcal{W}}
= j_{1\mu}\left(l_1,l'_1 \right)
\frac{i P^{\mu\mu'}(k_1)}{k_1^2 - m_1^2}
j_{2\nu}\left(l_2,l'_2 \right)
\frac{i P^{\nu\nu'}(k_2)}
{k_2^2 - m_2^2}
\mathcal{M}^{\mathcal{W}}_{\mu'\nu'} \, ,
\label{TotalAmp-1}
\end{eqnarray}
where $k_i= l_i -l'_i$ $(k_i^2\leq 0)$ 
and $P^{\alpha\beta}(k) = -g^{\alpha \beta}+
k^{\alpha} k^{\beta}/m_V^2$ for massive gauge-bosons in the unitary gauge.
The four-vectors $j_{1}^{\mu}$, $j_{2}^{\nu}$ denote
the  charged or neutral fermion currents and $\mathcal{M}^{\mathcal{W}}_{\mu'\nu'}$ 
is the vector-boson fusion amplitude for the process $V_1V_2\to \mathcal{W}$ that must be evaluated
 for off-shell gauge bosons.

Depending on whether the pairs $f_1, f'_1$ and $f_2, f'_2$ are particles or antiparticles, the
current  $j_{1}^{\mu}$ or $j_{2}^{\nu}$ is either composed of $u$ or $v$ Dirac spinors:
  \begin{equation}\label{eq:currentuv}
   e {\bar u}_{f'}(l')(a\gamma^\mu + b\gamma^\mu\gamma_5)u_f(l) \qquad \text{or}
    \qquad e {\bar v}_{f}(l)(a\gamma^\mu + b\gamma^\mu\gamma_5)v_{f'}(l') \, ,
  \end{equation}
 where $e$ denotes the positron charge.
 We are interested in the processes \eqref{The2To3Process} at high energies where the masses 
 of the light fermions $f_i, f'_i$ can be safely neglected, i.e., where $k_i^\mu j_{i\mu}=0$ $(i=1,2)$
 holds to very good approximation. 
 In order to decompose $g^{\mu\nu}$ we introduce two sets of
  polarization vectors $\varepsilon^{\mu}_j(\lambda)$ $(j=1,2)$
 that are mutually orthogonal and orthogonal to $k^{\mu}_j$
  and obey the normalization convention
   \begin{equation}\label{eq:normconeps}
    \varepsilon_j(\lambda)\cdot \varepsilon^*_j(\lambda')=(-1)^\lambda\delta_{\lambda,\lambda'} \, ,
    \qquad j=1,2 \,, \quad \lambda =0,\pm 1 \, .
   \end{equation}
An explicit representation of $\varepsilon^{\mu}_j(\lambda)$ in the center-of-mass frame of $V_1$ and $V_2$
 is given in appendix~A. With these polarization vectors one obtains
\begin{eqnarray}
-g^{\mu\nu} =
- \frac{k^{\mu}_j k^{\nu}_j}{k^2_j} +
\sum_{\lambda = \pm 1,0}
\left(-1\right)^{\lambda+1}
\varepsilon^{*\mu}_j(\lambda)
\varepsilon^{\nu}_j(\lambda) \, , \quad j=1,2 \, 
\label{eq:polcomp}
\end{eqnarray}
which holds for any space-like four-momentum $k_j^\mu$.

With \eqref{eq:polcomp} one can rewrite \eqref{TotalAmp-1}:
\begin{eqnarray}
\mathcal{M}_{f_1 f_2 \rightarrow f'_1 f'_2 \mathcal{W}}
= i^2 
\sum_{\lambda_1,\lambda_2 }
\left(-1\right)^{\lambda_1+ \lambda_2}
\frac{j_{1}(l_1,l'_1) \cdot \varepsilon^{*}_1(\lambda_1) }
{k_1^2 - m_1^2}
\frac{j_{2}(l_2,l'_2) \cdot \varepsilon^{*}_2(\lambda_2) }
{k_2^2 - m_2^2} 
\mathcal{M}^{\mathcal{W}}_{\lambda_1 \lambda_2} \, ,
\label{TotalAmp-2}
\end{eqnarray}
where the labels $\lambda_1$, $\lambda_2$ take the values $0,\pm1$
 and $\mathcal{M}^{\mathcal{W}}_{\lambda_1 \lambda_2}
\equiv \varepsilon^{\mu}_1 (\lambda_1)
\varepsilon^{\nu}_2(\lambda_2)
\mathcal{M}^{\mathcal{W}}_{\mu,\nu}$.

Squaring \eqref{TotalAmp-2} and averaging and summing over
 the spins (and colors, in the case of quarks) of the initial-state  and final-state fermions $f_i$, $f'_i$,
 one gets
\begin{eqnarray}
\overline{
|\mathcal{M}_{f_1 f_2 \rightarrow f'_1 f'_2 \mathcal{W}}|^2}
= 4 \sum_{\lambda_1,\lambda'_1,\lambda_2,\lambda'_2}
\left(-1\right)^{\lambda_1+ \lambda_2+ 
\lambda'_1+ \lambda'_2}
\frac{T_1\left(\lambda_1,\lambda'_1\right)}
{\left(k_1^2 - m_1^2\right)^2}
\frac{T_2\left(\lambda_2,\lambda'_2\right)}
{\left(k_2^2 - m_2^2\right)^2}
\mathcal{M}^{\mathcal{W}}_{\lambda_1 \lambda_2}
\mathcal{M}^{\mathcal{W} *}_{\lambda'_1 \lambda'_2}
\label{eq:MT1T2av}
\end{eqnarray}
with 
\begin{eqnarray} 
T_i\left(\lambda_i,\lambda'_i\right) =
\frac{1}{4}\sum
j_{i}(l_i,l'_i) \cdot \varepsilon^{*}_i(\lambda_i)
j_{i}^{*}(l_i,l'_i) \cdot \varepsilon_i(\lambda'_i) \, ,
\label{eq:ferm-tens}
\end{eqnarray}
 and the sum in \eqref{eq:ferm-tens} refers to fermion-spin summation.
 
In \cite{Kuss:1995yv} both the diagonal $(\lambda_i =\lambda'_i)$ and
non-diagonal $(\lambda_i \neq \lambda'_i)$ components of the helicity 
tensors \eqref{eq:ferm-tens} were taken into account in the computation
of the vector boson luminosities. The results of \cite{Kuss:1995yv} derived in the unitary gauge show that
the non-diagonal vector-boson luminosities are (significantly) smaller than the diagonal ones
if $0.2 \lesssim x < 1$, where 
\begin{equation}\label{eq:defxvar}   
x=\frac{(k_1+k_2)^2}{s} \equiv \frac{\hat s}{s} \, .
\end{equation}
 We consider in the following only the diagonal components of \eqref{eq:ferm-tens} 
 because i) the domain of applicability of the vector boson approximation is the region
where  $x$ is not very small and ii) because of the following conceptual issue. This approach
 loses its simplicity and appeal if the non-diagonal components are taken into account. Then the 
  resulting cross section can no longer be represented as in Eq.~\eqref{TotalCS-Final} below as 
  a sum of products of two-vector-boson luminosities times the 
   respective hard scattering $V_1 V_2$ cross sections.
 
In order to simplify the notation we use
$T_i(\lambda_i)\equiv e^{-2} T_i\left(\lambda_i,\lambda_i\right)$ from now on.
Furthermore, we define
\begin{eqnarray}
\tilde{\sigma}\left(\hat{s}, k_1^2, k_2^2; \lambda_1, \lambda_2 \right)
\equiv \frac{1}{2 \kappa_0 }
\int \mathrm{d} \tilde{p}_{\mathcal{W}}~
\left(2\pi\right)^4 
\delta^{(4)}\left(k_1+k_2 - p_{\mathcal{W}} \right)
|\mathcal{M}^{\mathcal{W}}_{\lambda_1 \lambda_2}|^2,
\label{eq:V1V2off}
\end{eqnarray}
where
\begin{eqnarray}
\kappa_0 &=& \sqrt{
\hat{s}^2 + m_1^4 + m_2^4  - 2 \hat{s}  m_1^2 - 2 \hat{s}  m_2^2 - 2 m_1^2 m_2^2}
\end{eqnarray}
 and $\hat{s}$  is the  squared
invariant mass  of the intermediate gauge-boson pair defined in \eqref{eq:defxvar}.   
Eq.~\eqref{eq:V1V2off} may be interpreted as the cross section for 
off-shell gauge-boson fusion  $V_1 V_2\to {\mathcal{W}}$, 
where the on-shell flux factor $\kappa_0$ is introduced by convention and for later convenience.
Using \eqref{eq:MT1T2av}, keeping only the diagonal contributions, and using
the definition \eqref{eq:V1V2off}, Eq.~\eqref{TotalCS-starting} becomes
\begin{eqnarray}
\sigma_{f_1 f_2 \rightarrow f'_1  f'_2 \mathcal{W}}
&=&
\left(\frac{\alpha}{\pi }\right)^2
\frac{4 \kappa_0 }{s}
\int \mathrm{d}\Gamma_2
\sum_{\lambda_1,\lambda_2}
\frac{1}
{\left(k_1^2 - m_1^2\right)^2}
\frac{1}
{\left(k_2^2 - m_2^2\right)^2}
\mathcal{L}_{\lambda_1 \lambda_2}~
\tilde{\sigma}(\lambda_1,\lambda_2) \, ,
\label{TotalCS-2}
\end{eqnarray}
where $\alpha=e^2/(4\pi)$ denotes the electromagnetic fine structure constant and  
\begin{eqnarray}
\mathcal{L}_{\lambda_1 \lambda_2} = T_1\left(\lambda_1\right)
T_2\left(\lambda_2\right) \, .
\label{FiveFoldLum}
\end{eqnarray} 
The helicity  tensors $T_i$ defined in  \eqref{eq:ferm-tens}, which are needed for computing 
the quantities $\mathcal{L}_{\lambda_1 \lambda_2}$, can be decomposed
as follows:   
\begin{equation}
T_i\left(\lambda_i\right) =
(v_i^2 + a_i^2) \mathcal{C}_i\left(\lambda_i\right)
+ 2 v_i a_i \mathcal{S}_i\left(\lambda_i\right) \, , \quad i=1,2 \, ,
\label{eq:Tidec}
\end{equation}
where     
\begin{eqnarray}
\mathcal{C}_j(\lambda_j)
&=&
\left(
l_j^{\mu}l_j^{'\nu} + l_j^{'\mu}l_j^{\nu}
- l_j \cdot l_j'~ g^{\mu\nu} 
\right) 
\varepsilon^{*}_{j\mu}(\lambda_j)
\varepsilon_{j\nu}(\lambda_j)  \, , \nonumber\\
\mathcal{S}_j(\lambda_j)
&=&
- i(-1)^{r_{j}}\epsilon_{\mu\nu\rho\sigma}
l_j^{'\mu} \varepsilon^{*\nu}_j(\lambda_j)
l_{j}^{\rho}\varepsilon^{\sigma}_j(\lambda_j) \, , \quad j=1,2 \, .
\label{CandSexpression}
\end{eqnarray}
Here we use the convention $\epsilon_{0123} = -1$ and $v_i$, $a_i$ are the vector and axial vector couplings
of the gauge boson $V_i$ in the parametrizations \eqref{eq:currentuv} of the currents.
For charged currents in the Standard Model they are given by $v_i=-a_i=1/(2\sqrt{2}\sin\theta_W)$, times the 
Cabibbo-Kobayashi-Maskawa mixing matrix element $V_{qq'}$ in the case of quarks.
The neutral current couplings are $v_i=(T_3^{f_i}-2\sin^2\theta_W)/(2\sin\theta_W\cos\theta_W)$
and $a_i= -T_3^{f_i}/(2\sin\theta_W\cos\theta_W)$. 

  In Eq.~\eqref{CandSexpression} the power $r_{j}=0$ $(r_{j}=1)$ if the label $j=1,2$
 refers to a particle (antiparticle) pair $f_j, f'_j$, i.e., the sign factor depends on whether 
the fermionic currents \eqref{eq:currentuv} involve $u$- or $v$-spinors.  \\

Rather than working with the nine quantities $\mathcal{L}_{\lambda_1 \lambda_2}$, 
it is convenient  to use in \eqref{TotalCS-2} the following linear combinations:
\begin{eqnarray}
\mathcal{L}_{\mathrm{TT}}
&\equiv& 
\mathcal{L}_{\mathrm{++}} +
\mathcal{L}_{\mathrm{+-}} +
\mathcal{L}_{\mathrm{-+}} +
\mathcal{L}_{\mathrm{--}}
= 4 (v_1^2 + a_1^2) (v_2^2 + a_2^2)
\mathcal{C}_1(\mathrm{+}) \mathcal{C}_2(\mathrm{+})
\, , \nonumber\\
\mathcal{L}_{\mathrm{T\overline{T}}}
&\equiv& 
\mathcal{L}_{\mathrm{++}} -
\mathcal{L}_{\mathrm{+-}} +
\mathcal{L}_{\mathrm{-+}} -
\mathcal{L}_{\mathrm{--}} 
= 8 (v_1^2 + a_1^2) (v_2 a_2)
\mathcal{C}_1(\mathrm{+}) \mathcal{S}_2(\mathrm{+})
\, ,\nonumber\\
\mathcal{L}_{\mathrm{\overline{T}T}}
&\equiv& 
\mathcal{L}_{\mathrm{++}} +
\mathcal{L}_{\mathrm{+-}} -
\mathcal{L}_{\mathrm{-+}} -
\mathcal{L}_{\mathrm{--}} 
= 8 (v_1 a_1) (v_2^2 + a_2^2)
\mathcal{S}_1(\mathrm{+}) \mathcal{C}_2(\mathrm{+})
\, ,\nonumber\\
\mathcal{L}_{\mathrm{\overline{T}\overline{T}}}
&\equiv& 
\mathcal{L}_{\mathrm{++}} -
\mathcal{L}_{\mathrm{+-}} -
\mathcal{L}_{\mathrm{-+}} +
\mathcal{L}_{\mathrm{--}} 
= 16 (v_1 a_1) (v_2 a_2)
\mathcal{S}_1(\mathrm{+}) \mathcal{S}_2(\mathrm{+})
\, ,\nonumber\\
\mathcal{L}_{\mathrm{TL}}
&\equiv& 
\mathcal{L}_{\mathrm{+0}} +
\mathcal{L}_{\mathrm{-0}} 
= 2 (v_1^2 + a_1^2) (v_2^2 + a_2^2)
\mathcal{C}_1(\mathrm{+}) \mathcal{C}_2(\mathrm{0})
\, , \nonumber\\
\mathcal{L}_{\mathrm{LT}}
&\equiv& 
\mathcal{L}_{\mathrm{0+}} +
\mathcal{L}_{\mathrm{0-}}
= 2 (v_1^2 + a_1^2) (v_2^2 + a_2^2)
\mathcal{C}_1(\mathrm{0}) \mathcal{C}_2(\mathrm{+})
 \, , \nonumber\\
\mathcal{L}_{\mathrm{\overline{T}L}}
&\equiv& 
\mathcal{L}_{\mathrm{+0}} -
\mathcal{L}_{\mathrm{-0}}  
= 4 (v_1 a_1) (v_2^2 + a_2^2)
\mathcal{S}_1(\mathrm{+}) \mathcal{C}_2(\mathrm{0})
\, ,\nonumber\\
\mathcal{L}_{\mathrm{L\overline{T}}}
&\equiv& 
\mathcal{L}_{\mathrm{0+}} -
\mathcal{L}_{\mathrm{0-}} 
= 4 (v_1^2 + a_1^2) (v_2 a_2) 
\mathcal{C}_1(\mathrm{0}) \mathcal{S}_2(\mathrm{+})
\, , \nonumber\\
\mathcal{L}_{\mathrm{LL}}
&\equiv& 
\mathcal{L}_{\mathrm{00}} 
=  (v_1^2 + a_1^2) (v_2^2 + a_2^2)
\mathcal{C}_1(\mathrm{0}) \mathcal{C}_2(\mathrm{0})
\, .
\label{FiveFoldLumRecomb}
\end{eqnarray}  

The $V_1V_2\to  \mathcal{W}$ cross sections  
$\tilde{\sigma}(\lambda_1,\lambda_2)$  in
\eqref{TotalCS-2} have to be transformed accordingly.   
One gets nine linear combinations in analogy to 
\eqref{FiveFoldLumRecomb}, but for each index T or ${\rm\overline{T}}$ an overall factor
$1/2$ is present. Thus, for instance,
\begin{eqnarray}   
\tilde{\sigma}_{\mathrm{TT}}  = \frac{1}{4}\left[
\tilde{\sigma}({\mathrm{+,+}}) +
\tilde{\sigma}({\mathrm{+,-}}) +
\tilde{\sigma}({\mathrm{-,+}}) +
\tilde{\sigma}({\mathrm{-,-}}) \right] \, , \nonumber \\
 \tilde{\sigma}_{\mathrm{\overline{T}L}}
   =     \frac{1}{2}   \left[
\tilde{\sigma}({\mathrm{+,0}}) -
\tilde{\sigma}({\mathrm{-,0}})    \right] \, , \quad 
\tilde{\sigma}_{\mathrm{LL}} =
\tilde{\sigma}({\mathrm{0,0}}) \, ,
\label{eq:tisireco}
\end{eqnarray}    
etc. Then \eqref{TotalCS-2} takes the form  
\begin{eqnarray}
\sigma_{f_1 f_2 \rightarrow f'_1  f'_2 \mathcal{W}}
&=&
\left(\frac{\alpha}{\pi }\right)^2
\frac{4 \kappa_0 }{s}
\int \mathrm{d}\Gamma_2
\sum_{\rm pol}
\frac{1}
{\left(k_1^2 - m_1^2\right)^2}
\frac{1}
{\left(k_2^2 - m_2^2\right)^2}
\mathcal{L}_{\rm pol}~
\tilde{\sigma}_{\rm pol} \, ,
\label{TotalCS-3}   
\end{eqnarray} 
where ``pol'' labels the nine polarization indices as in 
\eqref{FiveFoldLumRecomb}; i.e., 
${\rm pol}= \mathrm{TT},  \mathrm{T\overline{T}}$, etc.  
    
A basic issue of the effective vector boson method
is the modeling of the dependence of the off-shell
cross section $\tilde{\sigma}_{\rm pol}(\hat{s}, k_1^2, k_2^2)$ 
on $k_i^2$. If both $V_1$ and $V_2$ are transversely polarized, it turns out that
$\tilde{\sigma}_{\rm pol}$ is only slowly varying with $k^2_i$.  Thus, one can
put $\tilde{\sigma}_{\mathrm{TT}}(\hat{s}, k_1^2, k_2^2)
\simeq \tilde{\sigma}_{\mathrm{TT}}(\hat{s}, m_1^2, m_2^2)$   to good approximation.
If longitudinal polarizations are involved,   
$\tilde{\sigma}_{\rm pol}(\hat{s}, k_1^2, k_2^2)$ contains in the unitary gauge kinematic singularities
at $k^2_i=0$ which result from the longitudinal polarization vectors $\varepsilon^{\mu}_i(0)$.
The dependence on $k_i^2$ of $\varepsilon^{\mu}_i(0)$ (see \eqref{eq:polv1}, \eqref{eq:polv2}) suggests the following extrapolation \cite{Kuss:1995yv} 
of the on-shell $V_1V_2\to  \mathcal{W}$ cross sections  to off-shell values of $k_i^2$:
\begin{eqnarray}
\tilde{\sigma}_{\text{pol}}\left(\hat{s}, k_1^2,k_2^2\right)
\approx 
f_{\rm pol}\left(k_1^2,k_2^2\right)
 \hat{\sigma}_{\rm pol}
\left(\hat{s}, m_1^2, m_2^2 \right) \, , 
\label{eq:apprxcs}
\end{eqnarray}
 where $\hat{\sigma}_{\rm pol}$ is the on-shell $V_1 V_2\to {\cal W}$ cross section and 
\begin{eqnarray}
f_{\mathrm{TT}}
=  f_{\mathrm{T\overline{T}}}
= f_{\mathrm{\overline{T}T}} 
= f_{\mathrm{\overline{T}\overline{T}}} =1 \, ,  \nonumber \\
f_{\mathrm{TL}} =  f_{\mathrm{\overline{T}L}} = \frac{m_2^2}{-k_2^2} \, , 
\quad
f_{\mathrm{LT}} = f_{\mathrm{L\overline{T}}} = \frac{m_1^2}{-k_1^2} \, , \quad
f_{\mathrm{LL}} &=& \frac{m_1^2 m_2^2}{k_1^2 k^2_2} \, .
\label{EstimatingOffshell}
\end{eqnarray}

The quantities $\mathcal{L}_{\rm pol}$ defined in \eqref{FiveFoldLumRecomb}
are computed using 
\eqref{FiveFoldLum} -- \eqref{CandSexpression}. 
Because the helicities of a massive particle are dependent on the Lorentz frame, 
 we define the associated polarization vectors in the center-of-mass frame 
of $V_1$ and $V_2$, as already mentioned above.  They are given in Eqs.~\eqref{eq:polv1}, \eqref{eq:polv2}
of appendix~A.
The Minkowski scalar products which appear in the expressions for the form factors 
$C_1, S_1$ $(C_2, S_2)$ are conveniently evaluated in a Breit frame $B_1$ $(B_2)$
 which is defined such that only the $z$ component of the four-momentum $k_1^\mu$ $(k_2^\mu)$
is non-vanishing in  $B_1$ $(B_2)$. The  polarization vectors of $V_1$ $(V_2)$ defined in 
the $V_1 V_2$ center-of-mass frame 
must be Lorentz-transformed to $B_1$ $(B_2)$. 
The  resulting polarization vectors and four-momenta of $V_1$ $(V_2)$ 
in $B_1$ $(B_2)$ are given in  \cite{Kuss:1995yv}. 
We have computed the form factors $C_{1,2}$ and $ S_{1,2}$ using these parametrizations. 
Our  results agree with those given in appendix\footnote{The formula for 
$C_1(00)$ given in appendix B of  \cite{Kuss:1995yv} contains a misprint.
The sign in front of the third term in the square bracket should be positive.} B of \cite{Kuss:1995yv},
 up to an overall sign factor $(-1)^{r_j}$ associated with $S_i$. This factor appears if the 
  form factor $S_i$ is defined according to Eq.~\eqref{CandSexpression}.
 
It is appropriate to rewrite the phase-space integral 
in \eqref{TotalCS-3} in terms of new variables. One uses that $k_{1,2}^2<0$ in the physical region. 
Moreover, one uses  \eqref{eq:defxvar}   and 
\begin{equation}
u= 2 k_1\cdot l_2 + l_2^2 = 2 k_1\cdot l_2 \, ,
\label{eq:defxux}
\end{equation}
 and the azimuthal angle $\phi_1$ $(\phi_2)$ of the final-state
fermion $f'_1$ $(f'_2)$ in the Breit system $B_1$ $(B_2)$. 
 With \eqref{eq:apprxcs} the cross section \eqref{TotalCS-3} in the improved  effective vector boson approximation (IEVBA)
 in the unitary gauge takes the form
\begin{eqnarray}
\sigma^{\rm IEVBA}_{f_1 f_2 \rightarrow f'_1  f'_2 \mathcal{W}}
&=&
\sum_{\rm pol}
\int_{x_{\rm min}}^{1} \mathrm{d} x~ 
 \mathrm{\mathbf{L}}_{\rm pol} (x)
\hat{\sigma}_{\rm pol}
 \left(\hat{s} =x s, m_1^2, m_2^2\right) \, ,
\label{TotalCS-Final}
\end{eqnarray}
where
\begin{eqnarray}
\mathrm{\mathbf{L}}_{\rm pol}(x)
&\equiv& 
\left(\frac{\alpha}{2\pi }\right)^2
\frac{\kappa_0 }{s}
\int_{-s + \hat{s}}^{0} \mathrm{d} k_1^2 
\int_{-s + \hat{s}'}^{0} \mathrm{d} k_2^2
\int_{\hat{x} s}^{s} \frac{\mathrm{d} u }{ u}
\nonumber\\
&&~ ~ ~ ~ 
\times~\frac{k_1^2}
{\left(k_1^2 - m_1^2\right)^2}
\frac{k_2^2}
{\left(k_2^2 - m_2^2\right)^2}
f_{\rm pol} \mathcal{J}_{\rm pol}
\label{KS-Luminosity}
\end{eqnarray}
and 
\begin{eqnarray} 
\mathcal{J}_{\rm pol}
\equiv \frac{1}{k_1^2k_2^2}
\int_{0}^{2\pi} 
\frac{\mathrm{d} \phi_1}{ 2\pi}
\int_{0}^{2\pi} 
\frac{\mathrm{d} \phi_2}{ 2\pi}
\mathcal{L}_{\rm pol} \, .
\label{eq:JpolLpol}
\end{eqnarray}
The integration boundaries in \eqref{TotalCS-Final}
and \eqref{KS-Luminosity} are as follows: $x_{\rm min}={\hat{s}}_{\rm min}/s$,
where ${\hat{s}}_{\rm min} =p^2_{{\mathcal W}, \rm min}$ is the minimal value
of ${\hat{s}}$ for the production of the final state ${\mathcal W}$.
The variables $\hat{s}'$ and $\hat{x}$ which appear in the boundaries of the integrals in \eqref{KS-Luminosity}
are given by 
\begin{equation}
\hat{s}'= \frac{s}{s + k_1^2} \hat{s} \, , \qquad 
\hat{x} = \frac{1}{s}
\left(\nu + \frac{1}{2}\kappa\right)  \, ,
\label{eq:intbo}
\end{equation}
where 
\begin{equation}
 \nu=k_1 \cdot k_2 = \frac{1}{2}(\hat{s}- k_1^2 -k_2^2)\, , \quad \kappa = 2 \sqrt{\nu^2 -k_1^2 k_2^2} \, .
 \label{eq:defnukap}
\end{equation}

The dimensionless functions $\mathrm{\mathbf{L}}_{\rm pol}(x)$ are  the vector-boson pair luminosities
of $V_1$ and $V_2$. The product $\mathrm{\mathbf{L}}_{\rm pol}(x)dx$ can be interpreted
as the probability for emitting from $f_1$ and $f_2$ the vector bosons $V_1$ and $V_2$ with specified polarizations and with 
  squared $V_1 V_2$ center-of-mass energy in the interval $[xs,(x+dx)s]$. 
The nine functions $\mathcal{J}_{\rm pol}$ that depend, for fixed $f_1 f_2$ center-of-mass energy $\sqrt{s}$,  on the four variables
$k_1^2$, $k_2^2$, $x$ and $u$, are called differential luminosities. 
Our results for these functions are given in appendix~A.

Integrating the  $\mathcal{J}_{\rm pol}$ given in appendix~A with respect to $u$ 
(which can be done analytically), with the boundaries as in \eqref{KS-Luminosity}, 
we obtain three-fold differential luminosities.
For ${\rm pol}=\mathrm{TT}, \mathrm{LT}, \mathrm{TL}, \mathrm{LL}, \mathrm{\overline{T}\overline{T}}$ these
three-fold differential luminosities  were calculated before in \cite{Kuss:1995yv}.
 We agree with
the results\footnote{The formula for $J_{TL}$ in Eq.~40 of \cite{Kuss:1995yv} contains a 
misprint: the first term in the square bracket of the second line 
should read $3 s^2 \nu$.} of \cite{Kuss:1995yv} for ${\rm pol}=\mathrm{TT},\mathrm{LT},\mathrm{TL},\mathrm{LL}$,
 up to different normalization conventions used.
 The differential luminosity $\mathcal{J}_{\mathrm{\overline{T}\overline{T}}}$ originates from the product
 $\mathcal{S}_1(\mathrm{+}) \mathcal{S}_2(\mathrm{+})$ as Eq.~\eqref{FiveFoldLumRecomb} shows.
 If the fermion line $f_1,f_1'$ in Fig.~\ref{fig:WWscatt} refers to particles and $f_2,f_2'$ to antiparticles or vice
  versa, this product gets an overall factor $(-1)$ as explained below Eq.~\eqref{CandSexpression}.
 This distinction is not made in \cite{Kuss:1995yv} in the corresponding expression for $\mathcal{J}_{\mathrm{\overline{T}\overline{T}}}$.

Our  results for ${\rm pol}=\mathrm{\overline{T}T}, \mathrm{T\overline{T}},\mathrm{L\overline{T}},$ and $\mathrm{\overline{T}L}$
are not given  in \cite{Kuss:1995yv}. As mentioned in the introduction these luminosities are required if the matrix element 
$V_1(\lambda_1) V_2(\lambda_2)\to {\mathcal W}$
receives also contributions from parity-violating interactions, cf. Sec.~\ref{sec:applic}.

If one applies cuts on the rapidities of the particles in the final state then the 
integration range of $u$ is affected. Details are given in appendix~A. Thus in applications it is adequate to  perform  
this integration numerically, see Sec.~\ref{sec:applic}. ~\\
   
\begin{figure}[h!]
\centering
\includegraphics{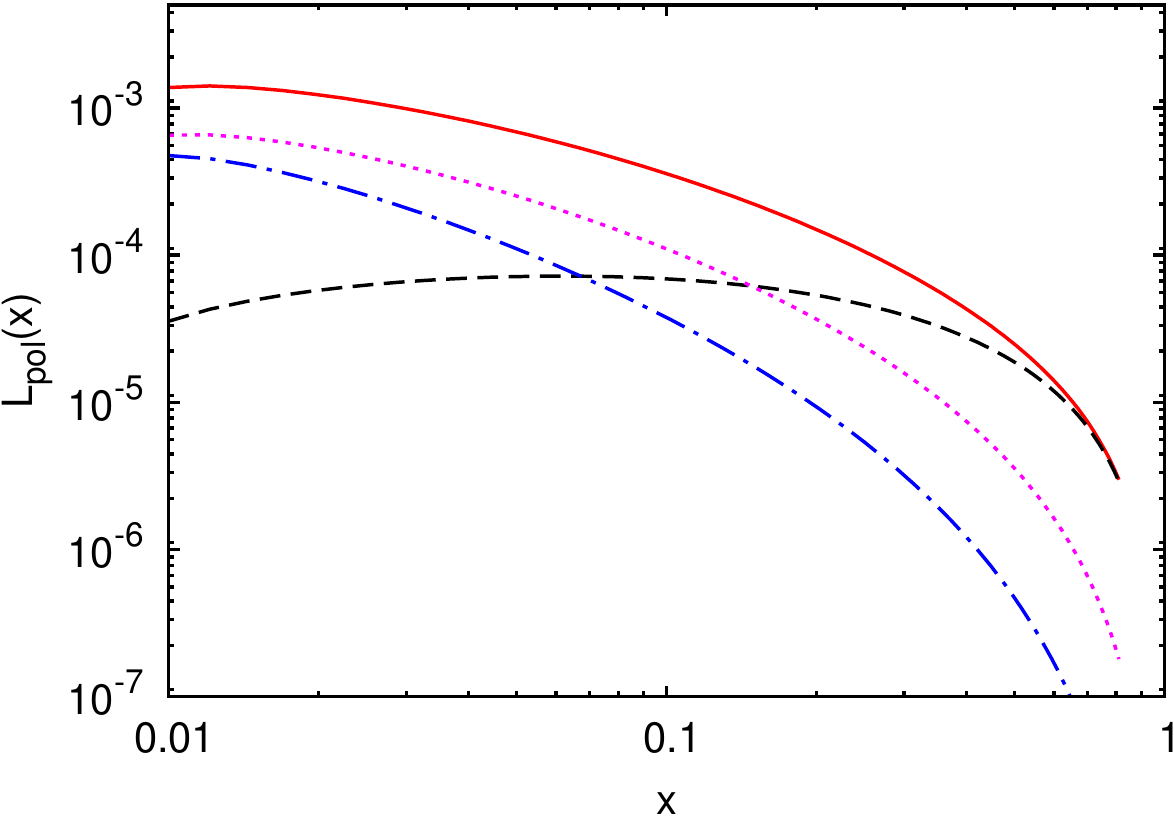}
\caption{The luminosities $\mathrm{\mathbf{L}}_{\rm T T}(x)$ (solid red), 
$\mathrm{\mathbf{L}}_{\rm L T}(x)$ (dotted magenta),
$\mathrm{\mathbf{L}}_{\rm L L}(x)$ (dot-dashed blue), and 
$(-1)\mathrm{\mathbf{L}}_{\rm \overline{T}\overline{T}}(x)$ (dashed black) in the unitary gauge for a $W^-W^+$ pair
in $e^-e^+$ collisions at $\sqrt{s}=2$~TeV.}
\label{fig:Lumi-xsec}
\end{figure}
 
 \begin{figure}[h!]
\centering
\includegraphics{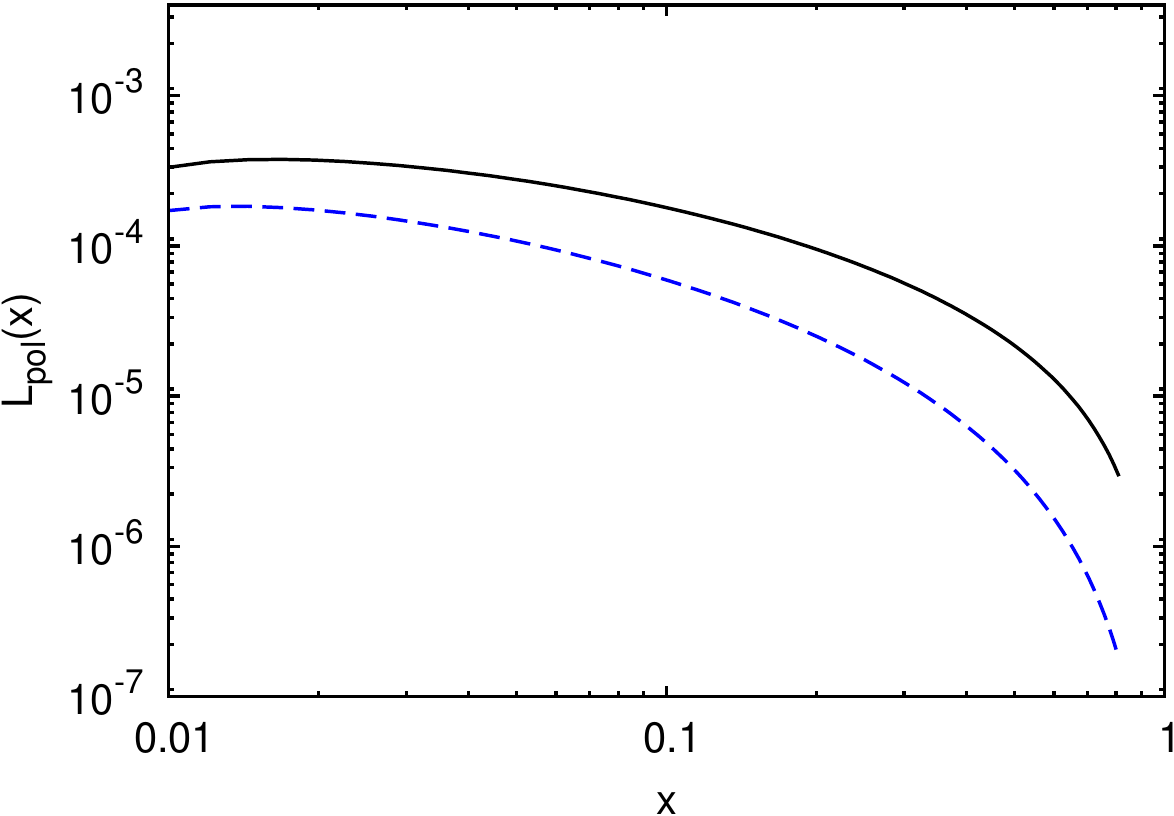} 
\caption{The luminosities $\mathrm{\mathbf{L}}_{\rm T \overline{T}}(x)$ (solid black) and
 $\mathrm{\mathbf{L}}_{\rm L \overline{T}}(x)$ 
(dashed blue)  in the unitary gauge for a $W^-W^+$ pair
 in $e^-e^+$ collisions at $\sqrt{s}=2$~TeV.}         
\label{fig:Lumi-pol}
\end{figure}

The differential luminosities    $\mathcal{J}_{\rm pol}$ given in appendix~A and the 
 formulae \eqref{TotalCS-Final} and \eqref{KS-Luminosity} apply to both 
quarks and leptons in the initial state.
In Fig.~\ref{fig:Lumi-xsec} and~\ref{fig:Lumi-pol} we show the luminosities 
$\mathrm{\mathbf{L}}_{\rm pol}(x)$ of finding 
a $W^-W^+$ pair in unpolarized $e^-e^+$ 
at $\sqrt{s}=2$ TeV.  For the computations of these luminosities 
we used $\alpha=1/137.035$, $m_W=80.385$ GeV $m_Z = 91.1876$ GeV and 
$\cos\theta_W=m_W/m_Z$.
Fig.~\ref{fig:Lumi-xsec} shows the cases   ${\rm pol}=\mathrm{TT}, \mathrm{LT},\mathrm{LL}$, 
and $\mathrm{\overline{T}\overline{T}}$. $CP$ invariance implies that the  luminosity 
$\mathrm{\mathbf{L}}_{\mathrm{TL}}=\mathrm{\mathbf{L}}_{\mathrm{LT}}$. 
The luminosities for ${\rm pol}=\mathrm{TT},\mathrm{LT},\mathrm{LL}$ 
agree with those displayed in Fig.~2 of \cite{Kuss:1995yv}.
Our luminosity for ${\rm pol}=\mathrm{\overline{T}\overline{T}}$, which is negative, differs from the corresponding one given in  \cite{Kuss:1995yv}
by an overall minus sign. This sign is convention-independent. The sign difference can be traced back to Eq.~\eqref{CandSexpression}. 
The form factor ${\cal S}_2(+)$ has a relative minus sign
compared with  ${\cal S}_1(+)$ because the incoming fermion $f_2=e^+$ is the antiparticle of $f_1$.

Fig.~\ref{fig:Lumi-xsec} shows that the luminosity $\mathrm{\mathbf{L}}_{\mathrm{TT}}$ for transversely 
polarized $W$ pairs is  the largest one.
  Needless to say, this does not  imply that the 
 contributions to \eqref{TotalCS-Final} from transversely polarized $W$ bosons  are always the dominant ones.

Fig.~\ref{fig:Lumi-pol} shows the luminosities  for
${\rm pol}=\mathrm{T\overline{T}}, \mathrm{L\overline{T}}$ that involve parity-odd combinations of vector and axial vector couplings.
The first (second) polarization index refers to the polarization of $W^-$ $(W^+)$ radiated from $e^-$ $(e^+)$. 
 These luminosities were not given in \cite{Kuss:1995yv}. For the example considered here, that is, $e^\mp\to W^\mp \nu_e/{\bar\nu_e}$, and
 for the case $q\to W^-q'$ and ${\bar q}\to W^+{\bar q}'$,  $CP$ invariance implies that
 \begin{equation} \label{eq:CPPVlumi}
 \mathrm{\mathbf{L}}_{\rm \overline{T}T}(x)= - \mathrm{\mathbf{L}}_{\rm T\overline{T}}(x) \, , \qquad
 \mathrm{\mathbf{L}}_{\rm \overline{T}L}(x) = - \mathrm{\mathbf{L}}_{\rm L \overline{T}}(x) \, .
 \end{equation}
 Relations between differential luminosities integrated with respect to $u$
  are given, for a general reaction \eqref{The2To3Process},  in Eq.~\eqref{eq:TLLTTTLT} of appendix~A.

If $V_1$ and/or $V_2$ is a $Z$ boson, the corresponding luminosities can be obtained 
in analogous fashion  by changing the value of the vector-boson mass $m_1$ and/or $m_2$, 
using the vector and axial-vector neutral current couplings given below \eqref{eq:Cpolcou}, and 
 by integrating $\mathcal{J}_{\rm pol}$. 
The  $V_1 V_2$ luminosities for vector bosons radiated off quarks are computed analogously.
    
 The above two-boson luminosities do not factorize into single boson distributions, because
in the above formulation, the emission of a gauge boson $V_1$ with definite helicity (defined in the $V_1 V_2$ center-of-mass frame) 
 from $f_1$ does depend on the 
squared off-shell mass $k_2^2$ of $V_2$, and vice versa. At high energies it seems justified to neglect this mutual 
dependence on $k_i^2$, because the fusion process is dominated by small momentum transfers.
Neglecting the dependence of the form factors $C_1, S_1$ $(C_2, S_2)$
on $k_2^2$ $(k_1^2)$ one obtains a luminosity formula $\mathrm{\mathbf{L}}^{\rm conv}_{\rm pol}(x)$ which can be represented
as a convolution of single vector boson distributions. 
These single $V$ distributions were first derived in \cite{Johnson:1987tj}.
A further approximation, the so-called  leading logarithmic 
approximation \cite{Cahn:1983ip,Dawson:1984gx,Kane:1984bb} (LLA), yields simplified expressions which have often been used in the literature.  
Here one performs the integral $\int \mathcal{J}_{\rm pol} du/u$ in \eqref{KS-Luminosity}  analytically.
 One neglects in the resulting expression the dependence on the $k_i^2$, performs the high-energy 
limit $s\ll m_i^2$, and keeps only the  leading logarithmic terms.
  
In this way, $\mathrm{\mathbf{L}}_{\rm pol} \to \mathrm{\mathbf{L}}^{\rm LLA}_{\rm pol}$.
These two approximations were analyzed in detail in \cite{Kuss:1995yv}. It was also shown by these authors
that the ratios $\mathrm{\mathbf{L}}^{\rm conv}_{\rm pol}/\mathrm{\mathbf{L}}_{\rm pol}$ are significantly larger than one
for almost all values of $x$; only for $x$ close to one, these ratios are also close to one.  Moreover, the ratio
$\mathrm{\mathbf{L}}^{\rm LLA}_{\rm pol}/\mathrm{\mathbf{L}}_{\rm pol}$ is even larger.
For $x\to 1$ this ratio is approximately close to one only for ${\rm pol}=\mathrm{LL}$ .
    
\section{The vector-boson pair luminosity in the axial gauge}
\label{sec:LumiAx}

In this section we derive the 
vector-boson pair luminosity in the axial gauge

Let us first recapitulate the salient features of the electroweak Standard Model in the axial gauge. 
The gauge-fixing term is chosen to be 
\begin{equation}\ \label{eq:axgf}
 {\cal L}_{gf} = -\frac{\xi}{2}\left[ (n\cdot A^a)^2 + (n\cdot B)^2 \right] \, ,
\end{equation}
 where $A^a_\mu$ and $B_\mu$ denote the SU(2)$_L$ und U(1)$_Y$ gauge fields and $n^\mu$ is a constant
  vector. As is well-known ghost fields are absent in this gauge, but the Goldstone fields are still present. We parametrize the SM Higgs doublet 
   field by $\Phi=(\phi_W, (v + H+i\phi_Z)/\sqrt{2})$, where $H$ is the physical Higgs field. The part of the Lagrangian
    bilinear in the gauge and Goldstone fields contains  terms that mix these fields. In order to proceed one  may either use propagators 
    that are non-diagonal in the  gauge-fields (cf.  \cite{Kunszt:1987tk}), or one diagonalizes these bilinear terms by  appropriate shifts
     of the Goldstone fields, as was done in \cite{Dams:2004vi}. As a consequence, the gauge and Goldstone fields decouple in the propagators, but
      the Feynman rules for the interactions vertices, given also in \cite{Dams:2004vi}, become more complicated than those in 
      the covariant  renormalizable gauges. We use the approach of  \cite{Dams:2004vi}. In this framework, the $W$ boson propagator
      is given in the limit $\xi \to\infty$ by 
   \begin{equation}\label{eq:axWprop}   
    iD^W_{\mu\nu}(k) = \frac{i  N_{\mu \nu}}{k^2 -m_W^2 +i\epsilon} \, , \qquad 
   N_{\mu\nu}(k)=  \left(-g_{\mu\nu}+\frac{n_\mu k_\nu + n_\nu k_\mu}{n\cdot k} - k_\mu k_\nu \frac{n^2}{(n\cdot k)^2} \right) \,.
   \end{equation}
The $Z$-boson (photon) propagator is obtained from \eqref{eq:axWprop}  by the replacement $m_W^2 \to m_Z^2$ $(m_W^2 \to 0)$. 

Because $N^{\mu\nu} n_\nu=0$ the symmetric propagator matrix $N^{\mu\nu}(k)$ has rank 3. Thus its spectral decomposition can be made in terms of three 
mutually orthogonal four-vectors $\varepsilon^{\mu}(\lambda)$, $\lambda =\pm1, 0$. We obtain, for any space-like four-momentum $k^\mu$:
\begin{equation} \label{axnmunu}
 N^{\mu\nu}(k) = \sum_{\lambda = \pm 1} \varepsilon^{*\mu}(\lambda)
\varepsilon^{\nu}(\lambda)  - \varepsilon^{\mu}(0)\varepsilon^{\nu}(0) \, ,
\end{equation}
where the dependence of the $\varepsilon^{\mu}$ on $k$ is not exhibited. The vectors that describe transverse polarization have to satisfy
\begin{equation*}
 k_\mu\varepsilon^{\mu} = n_\mu \varepsilon^{\mu} = 0 \, , \quad \varepsilon(\lambda)\cdot \varepsilon^*(\lambda')=\delta_{\lambda,\lambda'} \, ,
 \quad \lambda,\lambda' = \pm 1 \, .
\end{equation*}
Furthermore we get
\begin{equation} \label{axpol0}
\varepsilon^{\mu}(0) = \sqrt{\frac{-k^2}{(k\cdot n)^2 - n^2 k^2}}\left(n^\mu - \frac{n^2}{(k\cdot n)} k^\mu \right) \, .
\end{equation}

In the axial gauge the weak gauge-boson scattering amplitude depicted in Fig.~\ref{fig:WWscatt} is supplemented by diagrams where one or both of
 the propagators of the weak gauge bosons $V_1, V_2$ are replaced by the propagators of the Goldstone bosons $\phi_W, \phi_Z.$ However, because
  the couplings of  $\phi_W, \phi_Z$ to the fermions $f_i, f'_i$ $(i=1,2)$ are proportional to the fermion masses, these contributions vanish in 
  the limit $m_i, m'_i \to 0$, which we consider. Therefore, the  scattering amplitude analogous to \eqref{TotalAmp-1} is given  by
  \begin{eqnarray}
\mathcal{M}^{\rm axial}_{f_1 f_2 \rightarrow f'_1 f'_2 \mathcal{W}}
= j_{1\mu}\left(l_1,l'_1 \right)
\frac{i N^{\mu\mu'}(k_1)}{k_1^2 - m_1^2}
j_{2\nu}\left(l_2,l'_2 \right)
\frac{i N^{\nu\nu'}(k_2)}
{k_2^2 - m_2^2}
\mathcal{M}^{\mathcal{W}}_{\mu'\nu'} \, .
\label{AxTotalAmp-1}
\end{eqnarray}

 We decompose the two propagator matrices in \eqref{AxTotalAmp-1} according to \eqref{axnmunu}, \eqref{axpol0}. Then the matrix element
 \eqref{AxTotalAmp-1} takes the same form as the corresponding matrix element \eqref{TotalAmp-2}. Therefore the computation of the 
  cross section of $f_1 f_2 \rightarrow f'_1 f'_2 \mathcal{W}$ in the IEVBA in the axial gauge proceeds as the derivation in the unitary gauge in
  Sec.~\ref{sec:LumiF}. What is different now is the modeling of the relation between the off-shell and on-shell cross section for 
  $V_1 V_2 \to\mathcal{W}$. Because the longitudinal polarization vectors $\varepsilon_i^{\mu}(0)$ do not contain kinematic singularities at $k_i^2=0$
   we use, instead of \eqref{eq:apprxcs}, the approximation
\begin{eqnarray}
\tilde{\sigma}_{\text{pol}}^{\text{axial}}\left(\hat{s}, k_1^2,k_2^2\right)
\approx
 \hat{\sigma}_{\rm pol}
\left(\hat{s}, m_1^2, m_2^2 \right) \, , 
\label{ax:apprxcs}
\end{eqnarray}
 where $\hat{\sigma}_{\rm pol}$ is the on-shell $V_1 V_2\to {\cal W}$ cross section, which is gauge-independent. That is, we put all the factors
 $f_{\rm pol}=1$. 
  In our view, there is no physical argument for using in the axial gauge factors $f_{\rm pol}\neq 1$ in the extrapolation
     of the off-shell hard scattering cross section to the on-shell cross section, as done in the unitary gauge.

 The IEVBA approximation to the cross section of $f_1 f_2 \rightarrow f'_1 f'_2 \mathcal{W}$ is then given, 
 in analogy to \eqref{TotalCS-Final}, by
 \begin{eqnarray}
\sigma^{\rm IEVBA, axial}_{f_1 f_2 \rightarrow f'_1  f'_2 \mathcal{W}}
&=&
\sum_{\rm pol}
\int_{x_{\rm min}}^{1} \mathrm{d} x~ 
 \mathrm{\mathbf{L}}_{\rm pol}^{\text{axial}} (x)
\hat{\sigma}_{\rm pol}
 \left(\hat{s} =x s, m_1^2, m_2^2\right)  \, ,
\label{axTotalCS-Final}
\end{eqnarray}
where $\mathrm{\mathbf{L}}_{\rm pol}^{\text{axial}}(x)$ is obtained from \eqref{KS-Luminosity} using $f_{\rm pol}=1$ and 
$\mathcal{J}_{\rm pol}^{\text{axial}}$. In turn the differential luminosities $\mathcal{J}_{\rm pol}^{\text{axial}}$ are determined
 by the integral   \eqref{eq:JpolLpol} of  $\mathcal{L}_{\rm pol}^{\text{axial}}$. These quantities are defined as in \eqref{FiveFoldLumRecomb}
  with the form factors  $\mathcal{C}_i$,  $\mathcal{S}_i$ defined in \eqref{CandSexpression} to be computed in the axial gauge.
   One can choose the two sets of  transverse polarization vectors  $\varepsilon^{\mu}_i(\pm 1)$ to be identical to those in the unitary gauge
    if $n^\mu$ is appropriately chosen.
   Then only those (differential)
   luminosities change with respect to the corresponding ones in Sec.~\ref{sec:LumiF} where the label ``pol'' contains  at least one index L.
   We compute the axial-gauge form factors $\mathcal{C}_1(0)$ and $\mathcal{C}_2(0)$  in the Breit frames $B_1$ and $B_2$, respectively, which were 
   defined below \eqref{EstimatingOffshell}. For definiteness we choose in the following $n^\mu$ to be light-like, 
    and we use $n^\mu=(1,0,0,-1)$ in the $V_1 V_2$ center-of-mass frame.
   According to \cite{Accomando:2006mc} a light-like $n^\mu$ yields the best approximation to the cross-section ratio $\sigma^{\rm EVBA}/\sigma^{\rm full}$
    for $f_1 f_2 \rightarrow f'_1 f'_2 W^+W^-$.
    For this choice of $n^\mu$ the polarization vectors $\varepsilon^{\mu}_i(0)$ are given in Appendix~B in the $V_1 V_2$ center-of-mass frame and 
     in the frames $B_i$. Moreover, in this appendix we list also those $\mathcal{J}_{\rm pol}^{\text{axial}}$ which differ from their
      counterparts in the unitary gauge. In the following the term ``axial gauge'' refers to this choice of $n^\mu$.
    
\begin{figure}[h!]
\centering
\includegraphics{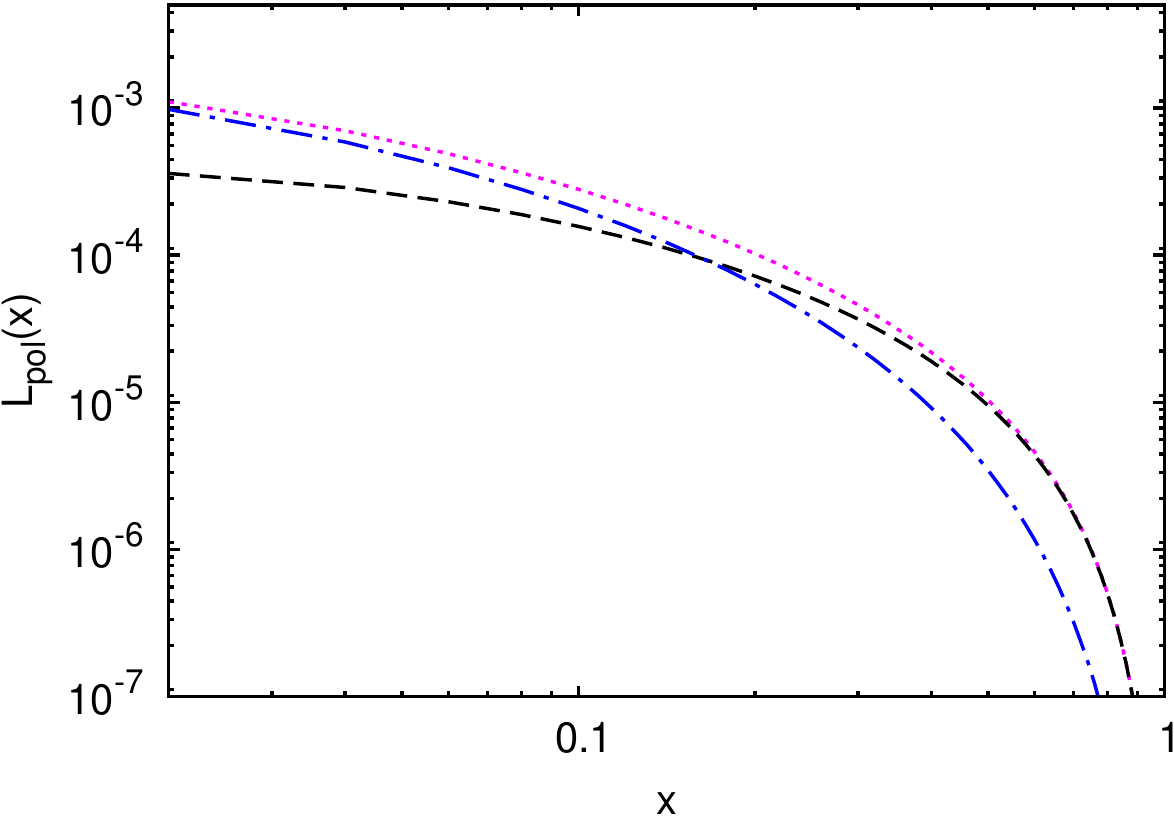}
\caption{The luminosities $\mathrm{\mathbf{L}}^{\text{axial}}_{\rm L L}(x)$ (dot-dashed blue), 
$\mathrm{\mathbf{L}}_{\rm L T}^{\text{axial}}(x)$ (dotted magenta), and 
 $\mathrm{\mathbf{L}}_{\rm L \overline{T}}^{\text{axial}}(x)$ (dashed black)  for a $W^-W^+$ pair
in $e^-e^+$ collisions at $\sqrt{s}=2$~TeV.}
\label{fig:axLumi-xsec}
\end{figure}

     Let us now consider, in analogy to Sec.~\ref{sec:LumiF}, the luminosities $\mathrm{\mathbf{L}}_{\rm pol}^{\text{axial}}(x)$
      of finding a $W^-W^+$ pair in unpolarized $e^-e^+$ collisions at $\sqrt{s}=2$ TeV. As mentioned above, 
      $\mathrm{\mathbf{L}}_{\rm T T}^{\text{axial}}$, $\mathrm{\mathbf{L}}_{\rm \overline{T}\overline{T}}^{\text{axial}}$,
      and $\mathrm{\mathbf{L}}_{\rm \overline{T}{T}}^{\text{axial}}= -\mathrm{\mathbf{L}}_{\rm {T}\overline{T}}^{\text{axial}}$ 
      are identical to those
      in the unitary gauge shown in Figs.~\ref{fig:Lumi-xsec} and~\ref{fig:Lumi-pol}. The other luminosities are 
      plotted in Fig.~\ref{fig:axLumi-xsec} where the same parameter values as in Sec.~\ref{sec:LumiF} were used.
      The relations \eqref{eq:CPPVlumi} hold also in the axial gauge.
      Moreover, $\mathrm{\mathbf{L}}_{\mathrm{TL}}^{\text{axial}}(x)=\mathrm{\mathbf{L}}_{\mathrm{LT}}^{\text{axial}}(x)$.
      
Comparing the luminosities displayed in Fig.~\ref{fig:axLumi-xsec} with the corresponding ones in 
Figs.~\ref{fig:Lumi-xsec} and~\ref{fig:Lumi-pol} we get the following. The luminosity 
$\mathrm{\mathbf{L}}^{\text{axial}}_{\rm L L}(x)$ is larger than $\mathrm{\mathbf{L}}_{\rm L L}(x)$
 by a factor $\sim 3$  for $x \sim 0.01 - 0.2$. The ratio of these two luminosities increases 
  to $\sim 7$   for $x\gtrsim 0.6.$ The luminosities $\mathrm{\mathbf{L}}_{\rm L T}^{\text{axial}}(x)$, 
 $\mathrm{\mathbf{L}}_{\rm L \overline{T}}^{\text{axial}}(x)$ are larger than the corresponding ones in the unitary gauge
  by a factor of $\sim 2 - 3.$  This is mainly due to the fact that in the axial gauge the factors \eqref{EstimatingOffshell}
  were not taken into account which suppress   the unitary-gauge luminosities in the region $|k_i^2|> m_i^2$.

\section{Applications and comparison with  full computations}
 \label{sec:applic}
   In this section we analyze the quality of the improved  effective vector boson approximation -- that is,
    the quality of     the formulae \eqref{TotalCS-Final} and \eqref{axTotalCS-Final} --       for
   the production cross section of $W^-W^+$ bosons and top-quark top antiquark $(\ttbar)$ pairs at high energies.
     To be specific we consider the processes $e^-e^+ \to W^-W^+ \nu_e \bar{\nu}_e$ and
   $e^-e^+ \to \ttbar + \nu_e \bar{\nu}_e$ in the Standard Model and compute the tree-level cross sections both 
    in the IEVBA using the weak-boson pair luminosities determined above in the unitary and axial gauge
    and fully, that is, taking all SM contributions into account, with the computer 
    code MadGraph \cite{Alwall:2014hca}.
   We determine the relative deviation of the IEVBA from the respective full cross section in dependence of several phase-space cuts.
   In both examples, non-diagonal interference contributions are not taken into account.
   As mentioned above, in our view the IEVBA loses its simplicity and appeal with these non-diagonal contributions.

    Besides the weak gauge-boson masses stated above, 
    we use $m_H=125$ GeV, $m_t=173$ GeV, and $m_b =4.7$ GeV for the Higgs-boson, top-quark, and $b$-quark mass, respectively.

\subsection{$e^-e^+ \rightarrow  W^-W^+ \nu_e \bar{\nu}_e$}
\label{susec:WW}
 We consider the reaction 
 \begin{equation} \label{eq:reacWW}
  e^-e^+ \rightarrow  W^-W^+ \nu_e \bar{\nu}_e
 \end{equation}
 for unpolarized $e^-e^+$ collisions  and center-of-mass energies $\sqrt{s}$ in the TeV range.
 At tree-level in the SM there are 56 diagrams that contribute to \eqref{eq:reacWW},
 while in the effective vector boson approximation 7 diagrams contribute to the hard scattering
  reactions $W^-W^+ \to W^-W^+$. 
   Within  the IEVBA the cross section for \eqref{eq:reacWW},
     summed over the helicities of the final-state $W^-W^+$, 
     is given in the unitary gauge by
\begin{equation} \label{eq:sigWWEWA}
\sigma_{W^-W^+}^{\rm{IEVBA}}(s)
=\sum_{\rm pol}
\int_{x_{\rm min}}^{1} \mathrm{d} x~ 
 \mathrm{\mathbf{L}}_{\rm pol} (x)
\hat{\sigma}^{W^-W^+}_{\rm pol}
 \left(\hat{s} =x s, m_W^2, m_W^2\right) \, ,
\end{equation}
where the sum extends over 
${\rm pol} = \mathrm{TT},  \mathrm{\overline{T}\overline{T}}, \mathrm{TL}, \mathrm{LT},  \mathrm{LL}$.
 An analogous formula holds in the axial gauge.
 Because at lowest order in the SM the scattering amplitude of $W^-W^+ \to W^-W^+$ is not affected by parity violation,
 the terms $\tilde{\sigma}^{W^-W^+}_{\rm pol}=0$ 
 for ${\rm pol}=\mathrm{\overline{T}T}, \mathrm{T\overline{T}},\mathrm{L\overline{T}},$ and $\mathrm{\overline{T}L}$.
  We define the  relative deviation of  \eqref{eq:sigWWEWA} from the full tree-level cross section $\sigma_{W^-W^+}^{\rm{full}}$
 computed with  MadGraph \cite{Alwall:2014hca} and the corresponding deviation in the axial gauge by
 \begin{equation} \label{eq:deltaWW}
  \delta_{WW} = \frac{\sigma_{W^-W^+}^{\rm{IEVBA}} -\sigma_{W^-W^+}^{\rm{full}}}{\sigma_{W^-W^+}^{\rm{full}}} \, ,
  \qquad 
  \delta_{WW}^{\rm axial} = \frac{\sigma_{W^-W^+}^{\rm{IEVBA, axial}} -\sigma_{W^-W^+}^{\rm{full}}}{\sigma_{W^-W^+}^{\rm{full}}} \, .
 \end{equation}
 In the following we choose $\sqrt{s}=2$ TeV. 
  The (improved) effective vector boson approximation is known to significantly overestimate the cross section for the reaction \eqref{eq:reacWW}
  unless appropriate cuts on kinematic variables of $W^\mp$ are made.
  We require a minimum value $M^*$ of the invariant mass $\mWW\equiv\hat s$ of the
  final-state $W^-W^+$ pair. First we analyze the quality of the IEVBA  for $W^-W^+$ production in the central region.
   We  compute, for fixed  $M^*$  the relative deviations $\delta_{WW}$ and $\delta_{WW}^{\rm axial}$
   for a sequence of upper cuts $y_W^*$
   on the moduli of the $W^-$- and $W^+$-boson rapidities in
 the laboratory frame; i.e, we restrict $|y_W|\leq y_W^*$. The implementation of this cut is described in appendix~A.
  The computation of the elastic $W^-W^+ \to W^-W^+$ cross section
  requires a cut in order to avoid the $t$-channel photon-propagator pole. Here we use a cut
    on the transverse momentum of the $W$ bosons, $p_{T,W}>20$ GeV. The same set of cuts is also applied to
     the calculation of $\sigma_{W^-W^+}^{\rm{full}}$. 
     
     The resulting values of $\delta_{WW}$  given in Table~\ref{tab:WWrapW}
    show that the size of the relative deviation depends quite sensitively on the rapidity cut. For loose cuts $y_W^*$  
     the cross section computed in the IEVBA approximation is larger than the exact value, while it is the other way around for very
      tight upper cuts on $|y_W|$. In the latter case the cross section is, however, reduced significantly. Table~\ref{tab:WWrapW}
       shows that the IEVBA approximation agrees within $\sim 10\%$ with the full calculation
        if $|y_W|$ is restricted to values less than $\sim 1.7$. The upper cut on $|y_W|$ can be losened if the cut  $M^*$  is increased.
        However, as the numbers in Table~\ref{tab:WWrapW} show, $|\delta_{WW}|$ increases again below $|y_W|=1.7$. 
        For $|y_W|<1.5$ the ratio $\delta_{WW}\simeq -.30$. For these tight cuts the event numbers rapidly decrease.
        
         As mentioned in Sec.~\ref{sec:LumiF} the
         luminosity $\mathrm{\mathbf{L}}_{\rm \overline{T}\overline{T}}(x)$ was given in  \cite{Kuss:1995yv} with the wrong sign.
         With the correct luminosities and with the  set of cuts used in Table~\ref{tab:WWrapW},  
         the approximation $\sigma_{W^-W^+}^{\rm{IEVBA}}$  improves by $1\%$ for $|y_W| <2.5$. 
          The improvement increases to $9\%$ for   $|y_W| \lesssim 1.7$.

\begin{table}[htbp]
\caption{Relative deviations $\delta_{WW}(\mWW\geq M^*)$ 
 and $\delta_{WW}^{\rm axial}(\mWW\geq M^*)$ 
defined in \eqref{eq:deltaWW} of the IEVBA cross section 
 from the full result for $e^-e^+\to W^-W^+ \nu{\bar\nu}$ at $\sqrt{s}=2$~TeV for several upper cuts $y_W^*$  on the moduli of
  the $W^\mp$-boson rapidities in
 the laboratory frame. The additional cut  $p_{T,W}>20$ GeV on the transverse momentum
  of the $W$ bosons was applied. \label{tab:WWrapW}}  
 \begin{center}
\begin{tabular}{c|cccccc}
 $y_W^*$                               & $2.5$  & $2$ &$1.8$ &$1.7$  & $1.6$   & $1.5$         \\ \hline 
$\delta_{WW}(\mWW\geq 400~{\rm GeV})$  & $3.05$ & $0.95$ &$0.33$ &$0.08$ &  $-0.12$ & $-0.26$  \\
 $\delta_{WW}(\mWW\geq 500~{\rm GeV})$ & $3.22$ & $0.71$ & $0.11$ &$-0.06$ &$-0.17$ & $-0.27$ \\                   
 $\delta_{WW}(\mWW\geq 600~{\rm GeV})$ & $3.20$ & $0.45$ &$0.05$ &$-0.07$ & $-0.18$ & $-0.26$  \\ \hline \hline
 $\delta_{WW}^{\rm axial}(\mWW\geq 400~{\rm GeV})$ & $5.15$ &$1.84$ &$0.86$  &$0.48$  &$0.17$  & $-0.03$   \\
 $\delta_{WW}^{\rm axial}(\mWW\geq 500~{\rm GeV})$ &$5.35$ & $1.38$ &$0.45$ & $0.21$ &$0.05$  & $-0.06$   \\
 $\delta_{WW}^{\rm axial}(\mWW\geq 600~{\rm GeV})$ &$5.22$ & $0.90$ &$0.34$ &$0.17$ & $0.04$ & $-0.06$   \\ \hline
\end{tabular}
\end{center}
\end{table}

 The corresponding ratios $\delta_{WW}^{\rm axial}$, which are given also in Table~\ref{tab:WWrapW}, show that for loose 
  upper  cuts on  $|y_W|$ the IEVBA approximation in the axial gauge is worse than in the unitary gauge.  This stems 
   from the fact that in the the axial gauge we have put all factors $f_{\rm pol}=1$  (cf. Sec.~\ref{sec:LumiAx}) which generates in this kinematic regime 
    larger   contributions to   $\sigma_{W^-W^+}^{\rm{IEVBA, axial}}$  with labels 
${\rm pol} = \mathrm{TL}, \mathrm{LT},  \mathrm{LL}$. Only for $|y_W| \lesssim 1.6$ the axial-gauge IEVBA provides a relatively good approximation to the 
 full cross section.
 
\begin{table}[htbp]
\caption{Same as Table~\ref{tab:WWrapW}, but $\delta_{WW}(\mWW\geq M^*)$ 
for several minimum cuts $p_{T,W}^*$  on the transverse momentum of the $W^\mp$ boson
 and the cut  $|y_W|\leq 2$.   \label{tab:WWptrans} }   
 \begin{center}
 \begin{tabular}{c|ccccc}
 $p_{T,W}^*$ [GeV]                     & $100$    & $150$  & $200$ &  $250$ & $300$ \\ \hline 
 $\delta_{WW}(\mWW\geq 400~{\rm GeV})$ & $0.359$ & $0.347$ & $0.271$ & $0.089$ & $-0.105$ \\
  $\delta_{WW}(\mWW\geq 500~{\rm GeV})$ & $0.440$ & $0.395$ & $0.312$ & $0.130$ & $-0.087$ \\
  $\delta_{WW}(\mWW\geq 600~{\rm GeV})$ & $0.488$ & $0.439$ & $0.315$ & $0.128$ & $-0.068$ \\ \hline \hline
 $\delta_{WW}^{\rm axial}(\mWW\geq 400~{\rm GeV})$ & $0.746$ &$0.684$ &$0.552$  &$0.314$  &$0.069$    \\ 
 $\delta_{WW}^{\rm axial}(\mWW\geq 500~{\rm GeV})$ & $0.843$ &$0.740$ &$0.603$  &$0.363$  &$0.089$    \\
 $\delta_{WW}^{\rm axial}(\mWW\geq 600~{\rm GeV})$ & $0.890$ &$0.786$ &$0.607$  &$0.362$  &$0.113$    \\ \hline
 \end{tabular}
\end{center}
\end{table}

Next we analyze $\delta_{WW}$ and $\delta_{WW}^{\rm axial}$ in dependence of a minimum cut on the transverse momentum of the $W^\mp$ boson.
 In addition a  cut  $|y_W|\leq 2$ on
   the $W^\mp$ rapidity is applied. 
 The results given in Table~\ref{tab:WWptrans} exhibit that the unitary-gauge IEVBA approximates the exact cross section
  to $\sim 10\%$ only if a cut $p_{T,W}\geq 250$ GeV is imposed. The additional  cut on
   the $W^\mp$ rapidity improves the quality of the IEVBA only for  $p_{T,W}\lesssim 200$ GeV.
  In the kinematic regime considered here the IEVBA in the axial gauge is in general worse than in the unitary gauge,
   for reasons mentioned above. Only for very hard cuts on $p_{T,W}$ the axial-gauge IEVBA works reasonably well.

\subsection{$e^-e^+ \rightarrow \ttbar  \nu_e \bar{\nu}_e$}

As a further reaction of interest, we investigate the cross section of 
  \begin{equation}\label{eq:reactt}
  e^-e^+ \rightarrow \ttbar  \nu_e \bar{\nu}_e 
 \end{equation}
 for unpolarized $e^-e^+$ collisions. 
 In the Standard Model twenty-one tree-level 
  Feynman diagrams contribute to \eqref{eq:reactt} while in the
   IEVBA the hard-scattering subprocess
   $W^-W^+\to \ttbar$ receives four diagram contributions. The cross section of \eqref{eq:reactt}
    in the IEVBA in the unitary gauge is 
\begin{equation} \label{eq:sigttEWA}
\sigma_{\ttbar}^{\rm{IEVBA}}(s)
=\sum_{\rm pol}
\int_{x_{\rm min}}^{1} \mathrm{d} x~ 
 \mathrm{\mathbf{L}}_{\rm pol} (x)
\hat{\sigma}^{\ttbar}_{\rm pol}
 \left(\hat{s} =x s, m_W^2, m_W^2\right) \, .
\end{equation}
 Here the sum extends over all nine polarization labels introduced in \eqref{FiveFoldLumRecomb}.
 That is, also the four luminosities and $\tilde{\sigma}^{\ttbar}_{\rm pol}$ that involve a parity-odd
  combination of vector and axial vector couplings contribute. This is because of the relations 
  \eqref{eq:CPPVlumi} and
 \begin{equation} \label{eq:CPsigtil}
\hat{\sigma}^{\ttbar} _{\rm \overline{T}T}(x)= -\hat{\sigma}^{\ttbar}_{\rm T\overline{T}}(x) \, , \qquad
 \hat{\sigma}^{\ttbar}_{\rm \overline{T}L}(x) = - \hat{\sigma}^{\ttbar}_{\rm L \overline{T}}(x) \, ,
 \end{equation}
 which follow from CP invariance. A formula analogous to \eqref{eq:sigttEWA} holds for the 
 IEVBA in the axial gauge.
 
 In analogy to \eqref{eq:deltaWW} we define the  relative deviation
 $\delta_{\ttbar}$ of \eqref{eq:sigttEWA} and the analogous ratio  $\delta_{\ttbar}^{\rm axial}$
 from the full tree-level cross section $\sigma_{\ttbar}^{\rm{full}}$ computed with  MadGraph.
  We choose $\sqrt{s}=2$ TeV and use the same set of minimum values $M^*$ as in Sec.~\ref{susec:WW},
   now for  the invariant mass $\mtt\equiv\hat s$ of the $\ttbar$ pair.
   
   First we analyze the quality of the IEVBA  in the unitary gauge.
   In Table~\ref{tab:ttrapt} the relative deviation $\delta_{\ttbar}$ is given 
   for a sequence of upper cuts $y_t^*$
   on the moduli of the $t$ and ${\bar t}$ rapidities in the laboratory frame.
   For rather loose cuts  a precision of  about $10\%$  or better can be obtained. However, similar to the example
    analyzed in subsection~\ref{susec:WW},
   the $y_t^*$ region where
      $\delta_{\ttbar}$ becomes minimal is correlated with the value of the cut on the $\ttbar$ invariant mass.
      We remark that the improvements discussed in Sec.~\ref{sec:LumiF} (i.e., correct sign of 
         $\mathrm{\mathbf{L}}_{\rm \overline{T}\overline{T}}$ and including the contributions to \eqref{eq:sigttEWA}
          with parity-odd  combinations of vector and axial vector couplings) improves the quality of the IEVBA
           by about $20\%$ $(30\%)$ for loose cuts ($y_t^*\lesssim 2$).

\begin{table}[htbp]
\caption{Relative deviations $\delta_{\ttbar}(\mtt\geq M^*)$ and $\delta_{\ttbar}^{\rm axial}$
 defined in analogy to \eqref{eq:deltaWW}  
 of the IEVBA cross section from the full result for 
$e^-e^+\to \ttbar \nu{\bar\nu}$ at $\sqrt{s}=2$~TeV for several upper cuts $y_t^*$  on the moduli of the 
 $t$ and $\bar t$ rapidities in the laboratory frame. \label{tab:ttrapt} }
\begin{center}
\begin{tabular}{c|cccccc}
 $y_t^*$  & $5$     & $4$     & $3$     &   $2$   & $1.5$   & $1$      \\ \hline 
 $\delta_{\ttbar}(\mtt\geq 400~{\rm GeV})$ & $0.090$ & $0.090$ & $0.090$ & $0.076$ & $0.011$ & $-0.081$  \\
 $\delta_{\ttbar}(\mtt\geq 500~{\rm GeV})$ & $0.064$ & $0.064$ & $0.064$ & $0.045$ & $-0.048$ & $-0.180$ \\                                           
 $\delta_{\ttbar}(\mtt\geq 600~{\rm GeV})$ & $0.006$ & $0.005$ & $0.004$ & $-0.024$ & $-0.154$& $-0.296$ \\ \hline \hline
 $\delta_{\ttbar}^{\rm axial}(\mtt\geq 400~{\rm GeV})$ & $3.18 $ & $3.18 $ & $3.17$ & $3.11$ & $2.78$ & $2.36$  \\
  $\delta_{\ttbar}^{\rm axial}(\mtt\geq 500~{\rm GeV})$ & $3.47$ & $3.47$ & $3.47$ & $3.38$ & $2.91$ & $2.31$  \\
  $\delta_{\ttbar}^{\rm axial}(\mtt\geq 600~{\rm GeV})$ & $3.55$ & $3.55$ & $3.55$ & $3.42$ & $2.74$ & $2.04$  \\ \hline 
\end{tabular}
\end{center}
\end{table}
  
 In addition, we analyze $\delta_{\ttbar}$ in dependence of a minimum cut $p_{T,t}^*$ on the transverse momentum
 of the $t$ and ${\bar t}$ quarks. The numbers given in Table \ref{tab:ttptrans} show that for $\ttbar$
 events with $\mtt\geq 500~{\rm GeV}$ and a moderate transverse momentum cut $p_{T,t}\geq 50$ GeV 
  a precision of about $10\%$ or better, depending on the value of $M^*$, can be obtained.

\begin{table}[htbp]
\caption{Same as Table~\ref{tab:ttrapt}, but $\delta_{\ttbar}(\mtt\geq M^*)$ 
for several minimum cuts $p_{T,t}^*$  on the $t$ and $\bar t$  transverse momentum. \label{tab:ttptrans}}  
 \begin{center}
 \begin{tabular}{c|ccccc}
 $p_{T,t}^*$ [GeV]   & $0$     & $50$    & $100$  & $150$ &  $200$  \\ \hline 
 $\delta_{\ttbar}(\mtt\geq 400~{\rm GeV})$ & $0.090$ & $0.108$ & $0.119$ & $0.027$ & $-0.099$ \\
  $\delta_{\ttbar}(\mtt\geq 500~{\rm GeV})$ & $0.064$ & $0.043$ & $0.005$ & $0.010$ & $-0.068$ \\
  $\delta_{\ttbar}(\mtt\geq 600~{\rm GeV})$ & $0.005$ &	$-0.043$ & $-0.110$ & $-0.144$ & $-0.163$ \\ \hline \hline 
  $\delta_{\ttbar}^{\rm axial}(\mtt\geq 400~{\rm GeV})$ & $3.18$ &	$3.27$ & $3.37$ & $3.20$ & $2.83$ \\
   $\delta_{\ttbar}^{\rm axial}(\mtt\geq 500~{\rm GeV})$ & $3.47$ &	$3.38$ & $3.21$ & $3.21$ & $2.97$ \\
   $\delta_{\ttbar}^{\rm axial}(\mtt\geq 600~{\rm GeV})$ & $3.55$ &	$3.33$ & $2.99$ & $2.80$ & $2.67$ \\ \hline 
\end{tabular}
\end{center}
\end{table}
        
The numbers for $\delta_{\ttbar}^{\rm axial}$   given in   Tables~\ref{tab:ttrapt} and~\ref{tab:ttptrans} 
 show that in the kinematic regimes considered the IEVBA approximation \eqref{eq:sigttEWA} in the axial gauge overestimates
  the full result by a factor of about 3 to 4. 
  The reason is that the on-shell hard scattering cross sections 
        $\hat{\sigma}^{\ttbar}_{\rm pol}$ are dominated by those where $W^-$ and/or $W^+$ is longitudinally polarized and the 
         associated axial-gauge luminosities $\mathrm{\mathbf{L}}_{\rm pol}$ are significantly larger than those in the unitary gauge. 
   If one chooses tighter cuts than those used in Tables~\ref{tab:ttrapt} and~\ref{tab:ttptrans} the deviations 
   $\delta_{\ttbar}^{\rm axial}$  diminish, but at the cost of rapidly decreasing event numbers.   \\
        
 Our results  for the reactions \eqref{eq:reacWW} and \eqref{eq:reactt} show that the unitary-gauge IEVBA provides a relatively good approximation  to the full
  cross section if hard cuts on  $|y_W|$ and $p_{T,W}$ are applied, while the  axial-gauge IEVBA is worse in general. 
   Given a specific choice of cuts 
   it is not possible to make a quantitative a priori estimate of the quality of the IEVBA.
   What could then be the use of the IEVBA -- in particular, in view of the fact that computer codes such as 
 those of \cite{Kilian:2007gr,Alwall:2014hca} allow to compute tree-level cross sections exactly? One potential application, which keeps the 
  computational effort at bay, is to calculate the tree-level cross section fully by taking into account all contributing Feynman diagrams but
   to implement the radiative corrections to the respective hard scattering process $V_1 V_2 \to {\cal W}$ using the IEVBA.
   In Ref.~\cite{Accomando:2006hq} this strategy was pursued with the unitary-gauge IEVBA for the reaction \eqref{eq:reacWW} 
  and it was argued that this leads to quantitatively satisfactory results.

\section{Summary and conclusions}
\label{sec:concl}
 We revisited the improved effective vector boson approximation \cite{Kuss:1995yv} in the unitary gauge that was designed to catch the essence of 
  weak gauge boson scattering $V_1 V_2 \to {\cal W}$ in high-energy $p p$ and $e^-e^+$ collisions with an improved precision
   compared to the EVBA in the leading logarithmic approximation.  We computed the correlated two-vector-boson luminosities $\mathrm{\mathbf{L}}_{\rm pol}(x)$ for $V_1$, $V_2$ being radiated off a massless quark or
     lepton $f_1$ and $f_2$, respectively, for the nine combinations of the transverse and longitudinal polarizations of $V_1$ and $V_2$. 
   We clarified a sign issue that appears in some of the $\mathrm{\mathbf{L}}_{\rm pol}(x)$. Our results for the parity-even luminosities $\mathrm{\mathbf{L}}_{\rm pol}(x)$
    agree with those of \cite{Kuss:1995yv}, up to a sign in the case of $\mathrm{\mathbf{L}}_{\rm \overline{T}\overline{T}}$.
    Our results for the four luminosities that involve a parity-odd combination of vector and axial vector couplings were, to our knowledge,
     so far not available in the literature. They are required if the hard scattering amplitude of $V_1 V_2 \to {\cal W}$ is affected also by parity-violating interactions.
     For instance, this is the case for $V_1 V_2 \to f f'$ where $f, f'$ are heavy quarks or leptons. We computed also the correlated 
     two-vector-boson luminosities $\mathrm{\mathbf{L}}_{\rm pol}^{\text{axial}}(x)$ in the axial gauge, using a specific vector $n^\nu$. 
   
     Furthermore, we studied the reactions $e^-e^+ \to W^- W^+  \nu_e \bar{\nu}_e$
     and $e^-e^+ \to \ttbar  \nu_e \bar{\nu}_e$ within the Standard Model for large $e^-e^+$ center-of-mass energies by computing the respective
      tree-level cross section using the IEVBA in the unitary and axial gauge 
       and comparing these approximations with the full SM cross section computed with MadGraph \cite{Alwall:2014hca}.
       Here, our aim was to probe the quality of the formulae \eqref{TotalCS-Final} and \eqref{axTotalCS-Final}.
       We found that  the IEVBA in the unitary gauge provides a relatively good approximation to the full cross section if hard cuts on the rapidities
        and transverse momenta of the $W^-, W^+$, respectively  $t$,${\bar t}$  in the final state are applied. In the case of $\ttbar$ 
         the inclusion of the  luminosities with parity-odd combinations of vector and axial vector couplings improves the quality of the IEVBA by 
         $20 - 30 \%$ depending on the chosen cuts. Using the axial-gauge luminosities the IEVBA becomes worse in general, for reasons discussed above. 
         
        The  applicability of the  (improved) effective vector boson approximation is certainly limited because, for a given high-energy reaction  
         and a choice of cuts, it seems not possible to quantify a priori the precision of the approximation. At best one may use the 
          IEVBA, which is gauge dependent, for a semi-quantitative 
          estimate of the effect of the hard scattering process $V_1 V_2 \to {\cal W}$. For instance, one may use it to estimate the effect of
           radiative corrections to this subprocess, as mentioned at the end of Sec.~\ref{sec:applic}. The IEVBA may also be useful if new physics effects are 
            considered and if the new physics effects on $V_1 V_2 \to {\cal W}$ are dominated by one or a few helicity combinations of the weak gauge bosons.


\subsubsection*{Acknowledgements}

We wish to thank Hubert Spiesberger  for discussions.  
Long Chen is supported by a scholarship from the China Scholarship Council (CSC).

\section*{A: Four-fold differential luminosities in the unitary gauge}
\label{sec:appendixA}

Here  we give explicit expressions for the nine 
differential luminosities $\mathcal{J}_{\rm pol}$ in the unitary gauge defined 
in \eqref{eq:JpolLpol}. They are calculated as follows. One starts 
with the center-of-mass frame of the off-shell vector bosons $V_1$ and $V_2$ whose
 four-momenta are given by
\begin{equation} \label{eq:4momv12}
k_1^\mu=(k_{01},0,0,k) \, , \qquad k_2^\mu=(k_{02},0,0,- k) \, .
 \end{equation}
In this frame the polarization vectors of $V_1$ and $V_2$ in the unitary gauge of helicity $\lambda_1$ and $\lambda_2$,
 respectively, are given in the Jacob-Wick phase conventions:
\begin{equation} \label{eq:polv1}
\varepsilon_1^\mu(\pm)=\frac{1}{\sqrt{2}}(0,\mp 1,-i,0) \, , \qquad \varepsilon_1^\mu(0)=\frac{1}{\sqrt{-k_1^2}}(k,0,0,k_{01}) \, ,
 \end{equation}
\begin{equation} \label{eq:polv2}
\varepsilon_2^\mu(\pm)=\frac{1}{\sqrt{2}}(0,\pm 1,-i,0) \, , \qquad \varepsilon_1^\mu(0)=\frac{1}{\sqrt{-k_2^2}}(-k,0,0,k_{02}) \, .
 \end{equation}
 As already mentioned below Eq.~\eqref{EstimatingOffshell} the four-momentum and polarization vectors 
  of $V_1$ $(V_2)$ and the four-momenta of $f_1, f'_1$ $(f_2, f'_2)$ are Lorentz-transformed into the Breit frame $B_1$ $(B_2)$ 
 where the form factors $\mathcal{C}_1, \mathcal{S}_1$ $(\mathcal{C}_2, \mathcal{S}_2)$  defined in 
 \eqref{CandSexpression}  are conveniently computed. They determine the $\mathcal{L}_{\rm pol}$ defined in \eqref{FiveFoldLumRecomb}.
  Performing the integration over the azimuthal
angles in \eqref{eq:JpolLpol} we obtain  the differential luminosities  $\mathcal{J}_{\rm pol}$.
   For the sake of brevity we omit details of the computation; they are given in \cite{Kuss:1995yv}.
 
For fixed  squared center-of-mass energy  $\sqrt{s}$ of the initial fermions $f_1, f_2$ 
the  $\mathcal{J}_{\rm pol}$ are functions of
$k_1^2$, $k_2^2$ and the variables $x$
and $u$ defined in \eqref{eq:defxvar}   and \eqref{eq:defxux}, respectively. We obtain for the reactions \eqref{The2To3Process}:
\begin{eqnarray}
\mathcal{J}_{\mathrm{TT}}
&=& c_{\mathrm{TT}}~
 \left(
1+ \frac{4\left(u -\nu\right)^2}{\kappa^2}\right) \left(\frac{1}{2} 
+ \frac{s(s - u)}{u^2} \right.
\nonumber\\ 
& & \left. + \frac{k_1^2 k_2^2}{\kappa^2 u^4}
\left(k_1^2 k_2^2 + u^2  - 2 u \nu\right)
\left(u^2 - 6 u s + 6 s^2 \right)
\right) \, , 
\label{eq:JTT}
\end{eqnarray}
\begin{eqnarray}
\mathcal{J}_{\mathrm{LT}}
&=& c_{\mathrm{LT}}~
\left( 1+ \frac{4 \left(u -\nu\right)^2 } {\kappa^2} \right)
\left(\frac{s (s - u) }{u^2} \right.
\nonumber\\ 
&& \left. + \frac{k_1^2 k_2^2}{\kappa^2 u^4}
\left(k_1^2 k_2^2 + u^2  - 2 u \nu\right)
\left(u^2 - 6 u s + 6 s^2
\right)
\right) \, ,
\label{eq:JLT}
\end{eqnarray}
\begin{eqnarray}
\mathcal{J}_{\mathrm{TL}}
&=& c_{\mathrm{TL}}
\left(
-1 + \frac{4 \left(u -\nu\right)^2}{\kappa^2} \right)
\left(
\frac{1}{2} +
\frac{s \left(s - u\right) } {u^2} \right.\nonumber\\ 
& & \left. + \frac{k_1^2 k_2^2}{\kappa^2 u^4}
\left(
k_1^2 k_2^2 + u^2  - 2 u \nu
\right) \left(
u^2 - 6 u s + 6 s^2
\right) \right) \, , 
\label{eq:JTL}
\end{eqnarray}
\begin{eqnarray}
\mathcal{J}_{\mathrm{\overline{T}\overline{T}}}
&=& (-1)^{r_1+r_2}\frac{4 c_{\mathrm{\overline{T}\overline{T}}}}{\kappa^2 u^2}
\left(u -\nu \right)
\left(k_1^2 k_2^2 - u \nu\right)
\left(u - 2 s\right) \, ,
\label{eq:JTbTb}
\end{eqnarray}
\begin{eqnarray}
\mathcal{J}_{\mathrm{\overline{T}T}}
&=& (-1)^{r_1} c_{\mathrm{\overline{T}T}}~
\left(1 + \frac{4 \left(u -\nu\right)^2}{\kappa^2}\right)
\frac{k_1^2 k_2^2 - u \nu }{\kappa u^2}\left(u - 2 s\right) \, ,
\label{eq:JTbT}
\end{eqnarray}
\begin{eqnarray}
\mathcal{J}_{\mathrm{T\overline{T}}}
&=& 
\left(-1\right)^{r_{2}}
4 c_{\mathrm{T\overline{T}}}
\frac{(u - \nu)}{\kappa}
\left(
\frac{1}{2}
+
\frac{s (s - u) }
{u^2} \right. \nonumber\\ 
 & &  \left. +\frac{k_1^2 k_2^2}{\kappa^2 u^4}
\left(k_1^2 k_2^2 + u^2  - 2 u \nu
\right) \left(u^2 - 6 u s + 6 s^2\right)\right) \, , 
\label{eq:JTTbar}
\end{eqnarray}
\begin{eqnarray}
\mathcal{J}_{\mathrm{\overline{T}L}}
&=&   (-1)^{r_1}   c_{\mathrm{\overline{T}L}}~
\left( -1 +
\frac{4 \left(u -\nu\right)^2}{\kappa^2}\right)
\frac{k_1^2 k_2^2 - u \nu }{\kappa u^2}
\left(u - 2 s \right) \, , 
\label{eq:JTbL}
\end{eqnarray}
\begin{eqnarray}
\mathcal{J}_{\mathrm{L\overline{T}}}
&=& 
\left(-1\right)^{r_{2}}
4 c_{\mathrm{L\overline{T}}}
\frac{(u - \nu)}{\kappa}
\left(\frac{s (s - u)}{u^2} \right.\nonumber\\ 
&& \left. + \frac{k_1^2 k_2^2}{\kappa^2 u^4}
\left(k_1^2 k_2^2 + u^2  - 2 u \nu\right)
\left(u^2 - 6 u s + 6 s^2\right) \right) \, ,
\label{eq:JLTbar}
\end{eqnarray}
\begin{eqnarray}
\mathcal{J}_{\mathrm{LL}}
&=& c_{\mathrm{LL}}~
\left( -1 + \frac{4 \left(u -\nu\right)^2}{\kappa^2} \right)
\left(\frac{s (s - u) }{u^2} \right.
\nonumber\\ 
&& \left. + \frac{k_1^2 k_2^2}{\kappa^2 u^4}
\left(k_1^2 k_2^2 + u^2 - 2 u \nu\right)
\left(u^2 - 6 u s + 6 s^2 \right)
\right) \, ,
\label{eq:JLL}
\end{eqnarray}
where the variables $\nu$ and $\kappa$ are given in \eqref{eq:defnukap} and the powers $r_1, r_2$, which are either zero or one,
are defined below Eq.~\eqref{CandSexpression}. In \eqref{eq:JTT} -- \eqref{eq:JLL}  we have used the abbreviations
\begin{eqnarray}
 c_{\mathrm{TT}} &= & c_{\mathrm{LT}} = c_{\mathrm{TL}} = c_{\mathrm{LL}}  = 
 \left(v_1^2 + a_1^2\right)\left(v_2^2 + a_2^2\right) \, , \nonumber \\
 c_{\mathrm{T\overline{T}}} & = & c_{\mathrm{L\overline{T}}} = 2 \left(v_1^2 + a_1^2\right) v_2 a_2 \, ,\nonumber  \\
 c_{\mathrm{\overline{T}T}} & = & c_{\mathrm{\overline{T}L}} =2 \left(v_2^2 + a_2^2 \right) v_1 a_1 \, , \nonumber \\
 c_{\mathrm{\overline{T}\overline{T}}} & = & 4~v_1 a_1 v_2 a_2  \, ,
 \label{eq:Cpolcou}
\end{eqnarray}
where $v_i, a_i$ are the vector and axial vector coupling of the intermediate gauge boson $V_i$
 which are defined below Eq.~\eqref{CandSexpression}. 
    
If no phase-space cuts are applied and if one integrates over the variable $u$ and defines 
${\tilde{\mathcal{J}}}_{\rm pol}=\int_{{\hat x}s}^s (du/u) \mathcal{J}_{\rm pol}$, then the following relations hold
in the physical region defined by 
 the  integration regions over the remaining phase-space variables in \eqref{TotalCS-Final}, \eqref{KS-Luminosity}:
\begin{equation}
{\tilde{\mathcal{J}}}_{\mathrm{TL}} = {\tilde{\mathcal{J}}}_{\mathrm{LT}} \, , \quad 
{\tilde{\mathcal{J}}}_{\mathrm{T\overline{T}}} = (-1)^{r_2-r_1} \frac{c_{\mathrm{T\overline{T}}}}{c_{\mathrm{\overline{T}T}}}~{\tilde{\mathcal{J}}}_{\mathrm{\overline{T}T}} \, , \quad 
{\tilde{\mathcal{J}}}_{\mathrm{L\overline{T}}} = (-1)^{r_2-r_1} \frac{c_{\mathrm{L\overline{T}}}}{c_{\mathrm{\overline{T}L}}}~{\tilde{\mathcal{J}}}_{\mathrm{\overline{T}L}} \, . 
\label{eq:TLLTTTLT}
\end{equation}

Finally, we describe how cuts can be applied on the rapidities of the particles in the final state ${\cal W}$ 
of the reactions \eqref{The2To3Process}.
 We introduce the variables
\begin{eqnarray}
\mathit{z} \equiv
\frac{u + k_1^2}{s}
= \frac{2 k_1\cdot l_2 + k_1^2}
{2 l_1\cdot l_2 } \, , \qquad 
K^2 \equiv  \frac{u + k_1^2}{u} k_2^2 \, . 
\label{eq: defzK}
\end{eqnarray}
In terms of these variables the three-dimensional integration measure in \eqref{KS-Luminosity}
 is
\begin{eqnarray}
\int_{-s +\hat{s}}^{0} \mathrm{d} k_1^2 
\int_{-s + \hat{s}'}^{0} \mathrm{d} k_2^2
\int_{\hat{x} s}^{s} 
\frac{\mathrm{d} u}{ u} 
=
\int_{x}^{1} 
\frac{\mathrm{d}
 \mathit{z}}{ \mathit{z}}
\int_{-s(1-\mathit{z})}^{0} 
\mathrm{d} k_1^2
\int_{-s(\mathit{z}-x)}^{0}
\mathrm{d} K^2 \, .
\label{eq:transmeas}
\end{eqnarray}
In the context of the effective vector boson approximation the dominant 
 kinematic configuration corresponds to the intermediate vector boson $V_1$ and $V_2$
  moving collinear to the $f_1 f_2$ beam axis.
   Then the variable $z$ defined in \eqref{eq: defzK} is approximately equal to the 
    longitudinal momentum fraction of $V_1$ with respect to $f_1$. 
    Analogously we denote by $z'$ the longitudinal momentum fraction of $V_2$ with respect to $f_2$.
    The longitudinal velocity of the intermediate 
vector-boson pair $V_1 V_2$ in the $f_1f_2$ center-of-mass frame is $\beta_{V_1 V_2}
 = (\mathit{z} - \mathit{z}')/(\mathit{z} + \mathit{z}'),$ and the rapidity of the pair is
 \begin{eqnarray}
y_{V_1 V_2}
&=& \frac{1}{2}
\mathrm{ln} \left( \frac{1 + \beta_{V_1 V_2}}{ 1 - \beta_{V_1 V_2}} \right) 
=  \frac{1}{2} \mathrm{ln} \left(\frac{\mathit{z}^2}{x} \right) \, .
\end{eqnarray}
We consider now a particle $F$ in the final state ${\cal W}$ 
of the reaction \eqref{The2To3Process}. (In the examples analyzed in Sec.~\ref{sec:applic}
 $F$ corresponds to a $W$ boson or an (anti)top quark.) The rapidity of $F$ in the $f_1 f_2$ center-of-mass
  frame is given by 
\begin{equation} \label{eq:rapFLcms}
y_F = y_{V_1 V_2} +  y'_F \, ,
\end{equation}
where $y'_F= (1/2)\mathrm{ln}\left[(E'_F+p'_{3F})/(E'_F-p'_{3F})\right]$ is the rapidity of $F$ in the 
$V_1 V_2$ center-of-mass frame. Cuts on $y_F$ can be implemented using \eqref{eq:transmeas} and \eqref{eq:rapFLcms}.

\section*{B: Four-fold differential luminosities in the axial gauge}
\label{sec:appendixB}

Here  we list explicit expressions for those 
differential luminosities that differ from their counterparts in the unitary gauge.
For definiteness, we choose $n^\mu$ to be light-like. In   the $V_1 V_2$ center-of-mass frame we use
 $n^\mu=(0,0,0,-1)$.
In this  frame the four-momenta of $V_1$ and $V_2$ are given by \eqref{eq:4momv12}
 and their transverse polarization vectors can be chosen to be those listed in  \eqref{eq:polv1}, \eqref{eq:polv2}.
 Using \eqref{axpol0} and $n^2=0$  the longitudinal polarization vectors in this frame are
\begin{equation} \label{Baxpol0}
\varepsilon^{\mu}_i(0) = \sqrt{\frac{-k_i^2}{(k_i\cdot n)^2}}~n^\mu \,, \quad i=1,2 \, .
\end{equation}
As was done in Appendix~A the four-momentum and polarization vectors 
  of $V_1$ $(V_2)$ and the four-momenta of $f_1, f'_1$ $(f_2, f'_2)$ are Lorentz-transformed into the Breit frame $B_1$ $(B_2)$.
  We obtain for the longitudinal polarization vectors of $V_1$ and $V_2$:
 \begin{eqnarray}
 {(\varepsilon_1^{B_1})}^{\mu}(0) & = &   h_1~(e_0, e_1, 0, e_3) \, ,  \nonumber \\
 h_1 & = &  \frac{2 \sqrt{-k_1^2}}{\kappa}\, , \nonumber \\
 e_0 & = & \frac{1}{\sqrt{-k_1^2}} \left(\nu - \frac{k_1^2 k_2^2}{u} \right)   \, , \nonumber \\
 e_1 & = &  \frac{\sqrt{-k_2^2}}{u} \sqrt{k_1^2 k_2^2+u(u-2\nu)}  \, , \nonumber \\
 e_3 & = &   - \frac{\kappa}{2\sqrt{-k_1^2}} \, ,  \nonumber \\
 (\varepsilon_2^{B_2})^{\mu}(0) & = & (1,0,0,1) \, .
 \label{Baxpol120}
\end{eqnarray} 
The variables $u, \nu$, and $\kappa$ are defined in \eqref{eq:defxux} and \eqref{eq:defnukap}.
The transverse polarization vectors of $V_1$ $(V_2)$ and the four-momenta of    $V_1, f_1, f'_1$ $(V_2, f_2, f'_2)$
 in $B_1$ $(B_2)$ are given in appendix~A of \cite{Kuss:1995yv}, which we do not reproduce here for the sake of brevity.

 With these momenta and polarization vectors one can compute the helicity tensors \eqref{eq:ferm-tens}
  and the associated form factors in the
 frames $B_1$ and $B_2$. Concerning the  form factors defined in \eqref{CandSexpression} one has the 
 following. The $\mathcal{C}_i(\lambda_i=\pm 1)$
  and $\mathcal{S}_i(\lambda_i=\pm 1)$ are identical to those in the unitary gauge. 
   The $\mathcal{S}_i(\lambda_i=0)$ are zero because the longitudinal polarization vectors are real vectors. Thus one has to compute 
   only those differential luminosities $\mathcal{J}_{\rm pol}^{\text{axial}}$
   defined by  \eqref{eq:JpolLpol} with $\mathcal{L}_{\rm pol} \to \mathcal{L}_{\rm pol}^{\text{axial}}$
   where the label ``pol'' contains at least one index L. We obtain
  
  \begin{eqnarray}
\mathcal{J}_{\mathrm{LL}}^{\text{axial}}
&=& c_{\mathrm{LL}}~
\frac{h_1^2 F}{4 u^4 k_1^2}
\left( 1 - \frac{4 \left(u -\nu\right)^2 } 
{\kappa^2} \right) \, ,
\label{eq:JLL-Axial}
\end{eqnarray}

\begin{eqnarray}
\mathcal{J}_{\mathrm{LT}}^{\text{axial}}
&=& c_{\mathrm{LT}}~
\frac{- h_1^2 F}{4 u^4 k_1^2}
\left( 1 + \frac{4 \left(u -\nu\right)^2 } 
{\kappa^2} \right) \, ,
\label{eq:JTbL-Axial}
\end{eqnarray}

 \begin{eqnarray}
\mathcal{J}_{\mathrm{L\overline{T}}}^{\text{axial}}
&=& \left(-1\right)^{r_{2}} c_{\mathrm{L\overline{T}}}~
\frac{h_1^2 F}{u^4 k_1^2}
\frac{4 \left(\nu - u \right)} 
{\kappa} \, ,
\label{eq:JLTb-Axial}
\end{eqnarray}
where 
\begin{eqnarray}
F& =  & 4 u^2 \nu^2 s (s - u)  
 + \left(k_1^2 k_2^2\right)^2  \left(u^2 - 6  u  s + 6 s^2\right) \nonumber \\
& & + k_1^2  k_2^2 u  \left(u^3 - 12 \nu s^2 
- 2 u^2 (\nu + s) + 2 u s (6 \nu + s)\right) \, ,
\end{eqnarray}
 and the couplings $c_{\rm pol}$ are defined in \eqref{eq:Cpolcou}.

Moreover, we find that 
\begin{equation} \label{eq:axTLunit}
 \mathcal{J}_{\mathrm{TL}}^{\text{axial}} = \mathcal{J}_{\mathrm{TL}} \, , \qquad 
 \mathcal{J}_{\mathrm{\overline{T}L}}^{\text{axial}} = \mathcal{J}_{\mathrm{\overline{T}L}}
\end{equation}
The integrands of these differential luminosities involves the form factor ${\cal C}_2(\lambda_2=0)$ that 
 happens to be identical in the axial and unitary gauge. 
 Notice, however, that the associated luminosities  $\mathrm{\mathbf{L}}_{\rm pol}^{\text{axial}}(x)$
  differ from those in the unitary gauge because in the axial gauge the 
  factors \eqref{EstimatingOffshell} are not taken into account.


\begin{thebibliography}{99}

\bibitem{Aad:2012tfa} 
  G.~Aad {\it et al.} [ATLAS Collaboration],
  Phys.\ Lett.\ B {\bf 716}, 1 (2013)
  [arXiv:1207.7214 [hep-ex]].
  
\bibitem{Chatrchyan:2012xdj} 
  S.~Chatrchyan {\it et al.} [CMS Collaboration],
  Phys.\ Lett.\ B {\bf 716}, 30 (2012)
  [arXiv:1207.7235 [hep-ex]].
  
 
\bibitem{Chanowitz:1985hj} 
  M.~S.~Chanowitz and M.~K.~Gaillard,
  Nucl.\ Phys.\ B {\bf 261}, 379 (1985).

\bibitem{Bagger:1995mk} 
  J.~Bagger, V.~D.~Barger, K.~m.~Cheung, J.~F.~Gunion, T.~Han, G.~A.~Ladinsky, R.~Rosenfeld and C.-P.~Yuan,
  Phys.\ Rev.\ D {\bf 52}, 3878 (1995)
  [hep-ph/9504426].
  
\bibitem{Englert:2008tn} 
  C.~Englert, B.~Jager, M.~Worek and D.~Zeppenfeld,
  Phys.\ Rev.\ D {\bf 80}, 035027 (2009)
  [arXiv:0810.4861 [hep-ph]].
  
\bibitem{Doroba:2012pd} 
  K.~Doroba, J.~Kalinowski, J.~Kuczmarski, S.~Pokorski, J.~Rosiek, M.~Szleper and S.~Tkaczyk,
  Phys.\ Rev.\ D {\bf 86}, 036011 (2012)
  [arXiv:1201.2768 [hep-ph]].
  
  
\bibitem{Cahn:1983ip} 
  R.~N.~Cahn and S.~Dawson,
  Phys.\ Lett.\ B {\bf 136}, 196 (1984)
  [Phys.\ Lett.\ B {\bf 138}, 464 (1984)].
  
\bibitem{Dawson:1984gx} 
  S.~Dawson,
  Nucl.\ Phys.\ B {\bf 249}, 42 (1985).
  
\bibitem{Kane:1984bb} 
  G.~L.~Kane, W.~W.~Repko and W.~B.~Rolnick,
  Phys.\ Lett.\ B {\bf 148}, 367 (1984).
  
\bibitem{Dawson:1988ai}
S.~Dawson,
Phys.\ Lett.\ B {\bf 217}, 347 (1989).
doi:10.1016/0370-2693(89)90879-4

  
\bibitem{Kunszt:1987tk} 
  Z.~Kunszt and D.~E.~Soper,
  Nucl.\ Phys.\ B {\bf 296}, 253 (1988).
 
   
\bibitem{Borel:2012by} 
  P.~Borel, R.~Franceschini, R.~Rattazzi and A.~Wulzer,
  JHEP {\bf 1206}, 122 (2012)
  [arXiv:1202.1904 [hep-ph]].
  
\bibitem{Kleiss:1986xp}
  R.~Kleiss and W.~J.~Stirling,
  Phys.\ Lett.\ B {\bf 182} (1986) 75.
  doi:10.1016/0370-2693(86)91081-6
%
\bibitem{Johnson:1987tj} 
  P.~W.~Johnson, F.~I.~Olness and W.~K.~Tung,
  Phys.\ Rev.\ D {\bf 36}, 291 (1987).
  
\bibitem{Kuss:1995yv} 
  I.~Kuss and H.~Spiesberger,
  Phys.\ Rev.\ D {\bf 53}, 6078 (1996)
  [hep-ph/9507204].


 
\bibitem{Willenbrock:1986cr} 
  S.~S.~D.~Willenbrock and D.~A.~Dicus,
  Phys.\ Rev.\ D {\bf 34}, 155 (1986).
  
\bibitem{Dawson:1986tc} 
  S.~Dawson and S.~S.~D.~Willenbrock,
  Nucl.\ Phys.\ B {\bf 284}, 449 (1987).
  
\bibitem{Kauffman:1989aq} 
  R.~P.~Kauffman,
  Phys.\ Rev.\ D {\bf 41}, 3343 (1990).
  
\bibitem{Larios:1997ey} 
  F.~Larios, T.~M.~P.~Tait and C.~P.~Yuan,
  Phys.\ Rev.\ D {\bf 57}, 3106 (1998)
  [hep-ph/9709316].
  
  
\bibitem{Gunion:1986gm} 
  J.~F.~Gunion, J.~Kalinowski and A.~Tofighi-Niaki,
  Phys.\ Rev.\ Lett.\  {\bf 57}, 2351 (1986).
  
\bibitem{Accomando:2006mc} 
  E.~Accomando, A.~Ballestrero, A.~Belhouari and E.~Maina,
  Phys.\ Rev.\ D {\bf 74}, 073010 (2006)
  [hep-ph/0608019].
  
   
\bibitem{Accomando:2006hq}
  E.~Accomando, A.~Denner and S.~Pozzorini,
  JHEP {\bf 0703}, 078 (2007)
  [hep-ph/0611289].
  
\bibitem{Bouayed:2007rt} 
  N.~Bouayed and F.~Boudjema,
  Phys.\ Rev.\ D {\bf 77}, 013004 (2008)
  [arXiv:0709.4388 [hep-ph]].
  
\bibitem{Alboteanu:2008my} 
  A.~Alboteanu, W.~Kilian and J.~Reuter,
  JHEP {\bf 0811}, 010 (2008)
  [arXiv:0806.4145 [hep-ph]].

  
  
 
  
  
\bibitem{Kilian:2007gr} 
  W.~Kilian, T.~Ohl and J.~Reuter,
  Eur.\ Phys.\ J.\ C {\bf 71}, 1742 (2011)
  [arXiv:0708.4233 [hep-ph]].
 
 
 
\bibitem{Alwall:2014hca} 
  J.~Alwall {\it et al.},
  JHEP {\bf 1407}, 079 (2014)
  [arXiv:1405.0301 [hep-ph]].
 

\bibitem{Arnold:2011wj} 
  K.~Arnold {\it et al.},
  arXiv:1107.4038 [hep-ph].
  
  
\bibitem{Baglio:2014uba} 
  J.~Baglio {\it et al.},
  arXiv:1404.3940 [hep-ph].
  

\bibitem{Brehmer:2014pka} 
  J.~Brehmer, J.~Jaeckel and T.~Plehn,
  Phys.\ Rev.\ D {\bf 90},  054023 (2014)
  [arXiv:1404.5951 [hep-ph]].
  
\bibitem{Kuss:1996ww} 
  I.~Kuss,
  Phys.\ Rev.\ D {\bf 55}, 7165 (1997)
  doi:10.1103/PhysRevD.55.7165
  [hep-ph/9608453].
  
\bibitem{Dams:2004vi} 
  C.~Dams and R.~Kleiss,
  Eur.\ Phys.\ J.\ C {\bf 34}, 419 (2004)
  doi:10.1140/epjc/s2004-01734-4
  [hep-ph/0401136].


  

 
\end{thebibliography}
\end{document}